\pdfoutput=1
\documentclass[iop]{emulateapj}
\newcommand{\kms}{${\rm km~s}^{-1}$}
\usepackage{amsmath}

\begin{document}

\title{Tracing the Orphan Stream to 55 kpc with RR Lyrae Stars}

\author{Branimir Sesar\altaffilmark{1,5}}
\author{Carl J.~Grillmair\altaffilmark{2}}
\author{Judith G.~Cohen\altaffilmark{1}}
\author{Eric C.~Bellm\altaffilmark{1}}
\author{Varun B.~Bhalerao\altaffilmark{3}}
\author{David Levitan\altaffilmark{1}}
\author{Russ R.~Laher\altaffilmark{2}}
\author{Eran O.~Ofek\altaffilmark{4}}
\author{Jason A.~Surace\altaffilmark{2}}
\author{Sumin Tang\altaffilmark{1}}
\author{Adam Waszczak\altaffilmark{1}}
\author{Shrinivas R.~Kulkarni\altaffilmark{1}}
\author{Thomas A.~Prince\altaffilmark{1}}

\altaffiltext{1}{Division of Physics, Mathematics and Astronomy, California
                 Institute of Technology, Pasadena, CA 91125,
                 USA\label{Caltech}}
\altaffiltext{2}{Spitzer Science Center, California Institute of Technology,
                 Pasadena, CA 91125, USA\label{Spitzer}}
\altaffiltext{3}{Inter-University Centre for Astronomy and Astrophysics, Post
                 Bag 4, Ganeshkhind, Pune 411 007, India}
\altaffiltext{4}{Benoziyo Center for Astrophysics, Weizmann Institute of
                 Science, 76100 Rehovot, Israel\label{Weizmann}}
\altaffiltext{5}{Corresponding author: bsesar@astro.caltech.edu\label{email}}

\begin{abstract}
We report positions, velocities and metallicities of 50 $ab$-type RR Lyrae
(RRab) stars observed in the vicinity of the Orphan stellar stream. Using about 
30 RRab stars classified as being likely members of the Orphan stream, we study 
the metallicity and the spatial extent of the stream. We find that RRab stars in
the Orphan stream have a wide range of metallicities, from -1.5 dex to -2.7 dex.
The average metallicity of the stream is -2.1 dex, identical to the value
obtained by \citet{new10} using blue horizontal branch stars. We find that the
most distant parts of the stream (40-50 kpc from the Sun) are about 0.3 dex more
metal-poor than the closer parts (within $\sim30$ kpc), suggesting a possible
metallicity gradient along the stream's length. We have extended the previous
studies and have mapped the stream up to 55 kpc from the Sun. Even after a
careful search, we did not identify any more distant RRab stars that could
plausibly be members of the Orphan stream. If confirmed with other tracers, this
result would indicate a detection of the end of the leading arm of the stream.
We have compared the distances of Orphan stream RRab stars with the best-fit
orbits obtained by \citet{new10}. We find that model 6 of \citet{new10} cannot
explain the distances of the most remote Orphan stream RRab stars, and conclude
that the best fit to distances of Orphan stream RRab stars and to the local
circular velocity is provided by potentials where the total mass of the Galaxy
within 60 kpc is $M_{60}\sim2.7\times10^{11}$ $M_\odot$, or about 60\% of the
mass found by previous studies. More extensive modelling that would consider
non-spherical potentials and the possibility of misalignment between the stream
and the orbit, is highly encouraged.
\end{abstract}


\section{Introduction}\label{introduction}

One of the fundamental goals of Galactic astronomy is the determination of the
Galaxy's gravitational potential, because knowledge of it is required in any
study of the dynamics or evolution of the Galaxy. An important tool in this
undertaking are stellar tidal streams, remnants of accreted Milky Way satellites
that were disrupted by tidal forces and stretched into filaments as they orbited
in the Galaxy's potential. The orbits of stars in these streams are sensitive to
the properties of the potential and thus allow us to constrain the potential
over the range of distances spanned by the streams. Over the past several years,
the GD-1 stream \citep{gd06}, the Sagittarius tidal streams \citep{maj03}, and
the Orphan stream \citep{gri06,bel07a} have been applied to this problem and
have been used to constrain the circular velocity at the Sun's radius
\citep{krh10}, the total mass within 60 kpc \citep{new10}, and the shape of the 
dark matter halo potential \citep{lm10}.

While each of the above streams is valuable in its own right, the Orphan stream
seems to have several properties that make it almost ideal for inference of the
Galactic potential. The stream is narrow ($\sim2\arcdeg$ wide), extends over
$60\arcdeg$ of the sky, and probes distances from 20 to 50 kpc from the Sun
\citep{new10}. The misalignment between the stream and the orbit may be small
enough that fitting an orbit directly to the stream has no significant
consequences for the inferred shape of the Galactic potential (\citealt{sb13},
but see also Section~\ref{conclusions}). Its orbital plane is quite different
from that of the Sagittarius streams, and thus the Orphan stream may be an
important element in our attempts to understand the shape and orientation of the
dark matter halo potential (the so-called ``halo conundrum'', \citealt{lm10}).
Because it spans distances from 20 to 50 kpc from the Sun, the Orphan stream can
be used to test new models of the Galactic potential that are based on the orbit
of the Sagittarius stream (e.g., the model suggested by \citet{vch13} in which
the shape of the dark matter halo potential transitions from oblate to triaxial 
at $\sim30$ kpc).

A state-of-the-art description of the Orphan stream and its orbit has been
presented by \citet{new10}. In their study, Newberg et al.~used imaging data
from the Sloan Digital Sky Survey (SDSS; \citealt{yor00}) and the spectroscopic
data from the Sloan Extension for Galactic Understanding and Exploration
(SEGUE; \citealt{yan09a}) to select main-sequence turn-off (MSTO) and blue
horizontal branch (BHB) stars, and with them to trace the density, kinematics,
and the metallicity of the stream. Using these stars, they were able to trace
the stream up to 46 kpc from the Sun, after which the stream blended with the
background and was no longer detectable in their data. In a more recent study,
\citet{cas13} used low-resolution spectroscopy to identify several K giants, but
these stars are associated with the closer part of the Orphan stream at 22 kpc
from the Sun.

It was quite unfortunate that the stream could not be traced to greater
distances, as it prevented \citet{new10} from excluding certain models of the
Galactic potential. As they stated in their Section 10.2, ``if we could follow
the Orphan Stream just a little farther out into the halo, we would have a much 
better power to determine the halo mass, since the distances to the stream for
each case diverge'' (also see their Figures~12 and~14).

In this paper, we use $ab$-type RR Lyrae (RRab) stars to trace the Orphan stream
in the northern Galactic hemisphere, with a focus on tracing it beyond the
46-kpc limit reached by \citet{new10}. The main goal of this work is to support 
future studies of the Galactic potential by providing a clean sample of Orphan
stream RR Lyrae stars with precise distances and radial velocities (better than 
5\% and 15 km s$^{-1}$, respectively). The secondary goal of this work is to use
the kinematics and positions of RR Lyrae stars associated with the Orphan stream
to see whether new measurements exclude any of the models previously considered
by \citet{new10}.

Compared to MSTO, BHB and K giant stars used in previous studies of the stream,
RRab stars (fundamental-mode pulsators) have three important advantages. First,
they are bright stars ($M_r=0.6$ mag at ${\rm [Fe/H]}=-1.5$ dex) that can be
detected at large distances (5-120 kpc for $14 < r < 21$). Second, they are
standard candles ($\sim5\%$ uncertainty in distance; see Section~\ref{targets}),
and third, they have distinct, saw-tooth shaped light curves which make them
easy to identify given multi-epoch observations (peak-to-peak amplitudes of
$r\sim1$ mag and periods of $\sim0.6$ days). RR Lyrae type $c$ (RRc,
first-overtone pulsators) stars have more sinusoidal light curves, which are
less distinct and more difficult to separate from, for example, contact binary
systems (e.g., see Figure~5 of \citealt{ses10}). Furthermore, RRc stars are less
numerous than RRab stars (RRab to RRc ratio is about 3:1). Due to these reasons,
RRc stars are not used in this work.

Thanks to their distinct light curves, pure and highly complete samples
($\gtrsim95\%$; \citealt{ses10}) of RRab stars can be selected, allowing one to
trace and efficiently\footnote{That is, no observing time is wasted on non-RR
Lyrae stars.} follow up with spectroscopy even quite diffuse and distant halo
substructures (e.g., the Pisces Overdensity and the Cancer moving groups;
\citealt{ses10, ses12}). In contrast, samples of BHB stars selected using
photometry only will be either incomplete or will have a non-negligible to
severe contamination by non-BHB objects\footnote{The best color-based selections
of BHB stars achieve $\sim75\%$ purity at $\sim50\%$ completeness
\citep{sir04,bel10,vic12}.}, resulting in unnecessary spectroscopic follow-up of
non-BHB objects. More relevant to the broader topic of tracing halo
substructures, the contamination in samples of intrinsically sparse tracers
(such as BHB or RR Lyrae stars), even as small as 25\%, can be problematic as it
may cause appearance of false halo substructures (for a detailed discussion of
this problem, see Section 2.4 of \citealt{ses13}).

The paper is organized as follows. In Section~\ref{data}, we briefly describe
the surveys and the catalogs from which the target RR Lyrae stars were selected,
and then proceed to describe the spectroscopic data, their reduction, and the
measurement of radial velocities and metallicities of observed RR Lyrae stars.
In Section~\ref{members}, we use metallicities and distances to identify likely
members of Orphan stream. Likely members are then used to study the metallicity
of the stream and its spatial extent in Sections~\ref{metallicity}
and~\ref{extent}. In Section~\ref{comparison}, we compare the positions and
velocities of likely Orphan stream members to best-fit orbits obtained by
\citet{new10}, and then present our conclusions in Section~\ref{conclusions}.

\section{Data}\label{data}

\subsection{Overview of CSS, LINEAR, and PTF Surveys}\label{overview}

The RRab stars used in this paper come from three synoptic surveys: the Catalina
Real-Time Sky Survey (CRTS), Lincoln Near Earth Asteroid Research survey
(LINEAR), and the Palomar Transient Factory (PTF). In the remainder of this
section, we summarize each of these surveys, including an overview of the survey
parameters, and details on the quality of their photometry.

\begin{figure*}[!ht]
\epsscale{0.8}
\plotone{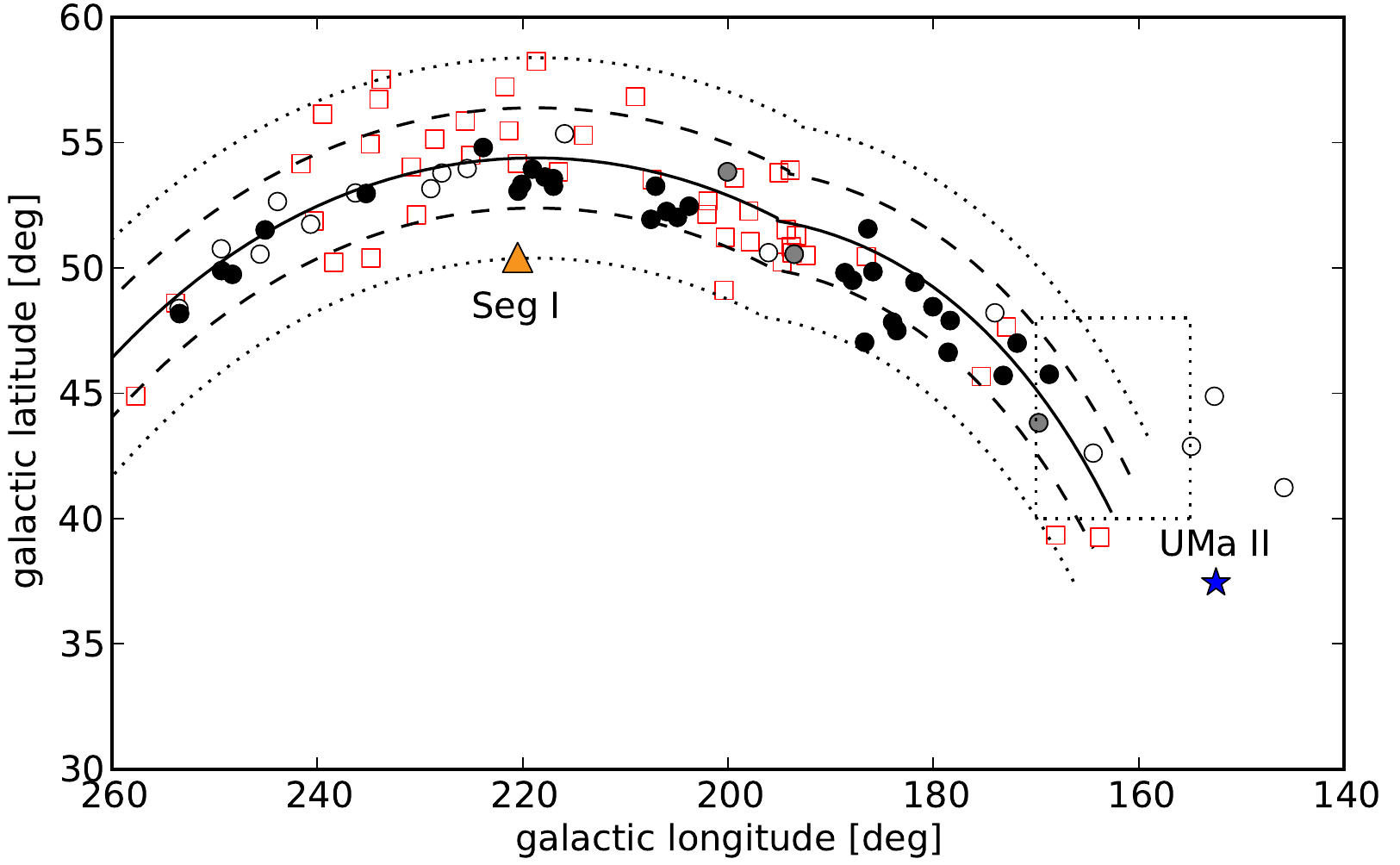}
\caption{
Spatial distribution of RRab stars in the vicinity of the Orphan stream. The
circles show stars with spectroscopic observations and the greyscale indicates
the likelihood of Orphan stream membership based on the star's velocity and
heliocentric distance (black is high, gray is medium, and white is low
likelihood; see Section~\ref{members}). The open squares show positions of CSS
RRab stars without spectroscopic observations located at heliocentric distances
greater than 30 kpc, but within $4\arcdeg$ of the Orphan stream equator
($|B_{Orphan}| < 4\arcdeg$) or within the region marked by the dotted box.
Spectroscopic followup may be desirable for some of these stars, especially for
those close to the predicted orbits of the Orphan stream (see the top left panel
of Figure~\ref{fig2}). The dotted box shows the region additionally searched for
distant RR Lyrae stars in PTF (see Section~\ref{extent}). The solid, dashed, and
dotted lines show Orphan stream latitudes ($B_{Orphan}$) of $0\arcdeg$,
$\pm2\arcdeg$, and $\pm4\arcdeg$, respectively. The ``kink'' in lines at
$l\sim195\arcdeg$ is due to corrected $B_{Orphan}$ for
$\Lambda_{Orphan}<-15\arcdeg$ ($l\lesssim195\arcdeg$), as defined by
Equation~\ref{B_corrected} (also see Section 4 of \citealt{new10}). For
comparison, the orange triangle and the blue star symbols show the locations of 
the Segue I \citep{bel07b} and Ursa Major II \citep{zuc06} dwarf spheroidal
(dSph) galaxies, respectively.
\label{fig1}} 
\end{figure*}

The CRTS\footnote{\url{http://crts.caltech.edu}} \citep{dra09} uses three
separate telescopes: the Catalina Sky Survey 0.7m Schmidt (CSS), the Mount
Lemmon Survey 1.5m (MLS), and the Siding Spring Survey 0.5m Schmidt (SSS). The
fields of view are, respectively, 8.1 deg$^2$, 1.2 deg$^2$, and 4.2 deg$^2$,
with corresponding limiting magnitudes 19.5, 21.5, and 19.0 mag (see Table 1 of
\citealt{lar07}). The magnitude zero-point of CRTS photometry is tied to the
Johnson $V$-band even though the observations are taken through a clear
filter\footnote{The filter response of the CSS clear filter is closer to Johnson
$R$-band than to Johnson $V$-band (A.~Drake 2013, private communication).}. The 
majority of data currently available is from CSS \citep{dra13}, and has a
typical cadence of one set of four exposures per night per field separated by 10
min, repeated every two weeks. The uncertainty in CRTS photometry is $\sim0.03$ 
mag for $V<16$ mag and rises to 0.1 mag at $V\sim19.2$ mag (see Figure~3 of
\citealt{dra13}).

The Lincoln Near Earth Asteroid Research survey\footnote{Public access to LINEAR
data is provided through the SkyDOT web site
(\url{https://astroweb.lanl.gov/lineardb}).} (LINEAR; \citealt{sto00}) uses two
telescopes at the White Sands Missile Range for a synoptic survey primarily
targeted at the discovery of near Earth objects. Each exposure covers $\sim2$
deg$^2$ to a $5\sigma$ limiting magnitude of $r\sim18$. \citet{ses11}
re-calibrated the LINEAR data using the SDSS survey, resulting in 200 unfiltered
observations per object (600 observations for objects within $\pm10\arcdeg$ of
the Ecliptic plane) for 25 million objects in the 9,000 deg$^2$ of sky where the
LINEAR and SDSS surveys overlap (roughly, the SDSS Galactic cap north of
galactic latitude 30 and the SDSS Stripe 82 region). The photometric precision
of LINEAR photometry is 0.03 mag at the bright end ($r\sim14$) and 0.2 mag at
$r = 18$ mag.

The PTF\footnote{\url{http://www.ptf.caltech.edu}} \citep{law09,rau09} is a
synoptic survey designed to explore the transient sky. The project utilizes the
48-inch Samuel Oschin Schmidt Telescope on Mount Palomar. Each PTF image covers 
7.26 deg$^2$ with a pixel scale of $1.01\arcsec$. The typical PTF cadence
consists of two 60-sec exposures separated by $\sim1$ hour and repeated every
one to five days. By June 2013, PTF observed $\sim11,000$ deg$^2$ of sky at
least 25 times in the Mould-$R$ filter\footnote{The Mould-$R$ filter is similar 
in shape to the SDSS $r$-band filter, but shifted {27 \AA} redward.} (hereafter,
the $R$-band filter), and about 2200 deg$^2$ in the SDSS $g^\prime$ filter. PTF
photometry is calibrated to an accuracy of about 0.02 mag \citep{ofe12a,ofe12b} 
and light curves have relative precision of better than 10 mmag at the bright
end, and about 0.2 mag at the survey limiting magnitude of $R=20.6$ mag. The
relative photometry algorithm is described in
\citet[see their Appendix A]{ofe11}.

To verify that the measurements taken by different surveys have the same
magnitude zero-point, we have compared median PTF, LINEAR, and CSS magnitudes
with SDSS $r$-band magnitudes. This comparison was done for non-variable sources
with the SDSS $g-r$ color in the 0.2-0.3 mag range (average $g-r$ color of RR
Lyrae stars). We found that the magnitudes of sources in this color range are
offset by $\Delta=-0.05$ mag (PTF), $\Delta=0.05$ mag (LINEAR), and
$\Delta=-0.05$ mag (CSS). 

\subsection{Spectroscopic Targets}\label{targets}

In Figure~\ref{fig1}, we show the spatial distribution of RRab stars targeted
for spectroscopy. The targets were taken from CSS \citep{dra13}, LINEAR
\citep{ses13}, and PTF catalogs of RR Lyrae stars (Sesar et al., in prep), and
are located in the vicinity of the Orphan stream.

Throughout this work we use the longitude $\Lambda_{Orphan}$ and latitude
$B_{Orphan}$ defined in the coordinate system of the Orphan stream by
\citet[see their Section 2]{new10}. As described in their Section 4, for
$\Lambda_{Orphan}< -15\arcdeg$ the Orphan latitude needs to be redefined as
\begin{equation}
B_{corr} = B_{Orphan} + 0.00628\Lambda^2_{Orphan} + 0.42\Lambda_{Orphan} + 5.
\label{B_corrected}
\end{equation}
To allow reproduction of coordinates used in this work, we provide (as
supplementary data) two functions written in Python that convert galactic to
Orphan coordinates and vice versa.

Briefly, to select RR Lyrae stars from PTF we first searched for variable PTF
sources that have SDSS colors consistent with colors of RR Lyrae stars
(Equations 6 to 9 of \citealt{ses10}). A period-finding algorithm was then
applied to light curves of color-selected objects, and objects with periods in
the range 0.2-0.9 days were kept. Light curves were then phased (period-folded)
and SDSS $r$-band RR Lyrae light curve templates constructed by \citet{ses10}
were fitted to phased data. Finally, we visually inspected template-fitted light
curves to tag sources as RR Lyrae stars. A more detailed description of the
selection of PTF RR Lyrae stars can be found in Section 3 of \citet{ses12}.

The positions and light curve parameters of RR Lyrae stars observed in this
work are listed in Table~\ref{table-positions}. The heliocentric distances were
calculated using the original flux-averaged magnitude,
$\langle m\rangle_{orig}$, corrected for the ISM extinction ($rExt$) and the
magnitude offset with respect to the SDSS $r$-band ($\Delta$)
\begin{equation}
\langle m \rangle = \langle m \rangle_{orig} - rExt - \Delta,
\end{equation}
where $rExt$ is the extinction in the SDSS $r$-band calculated using the
\citet{SFD98} dust map, and $\Delta$ is -0.05 mag (PTF), 0.05 mag (LINEAR), and
-0.05 mag (CSS; see the last paragraph in Section~\ref{overview}).

The absolute magnitude of an RRab star, $M_{RR}$ \citep{chaboyer99,cc03}, was
calculated as
\begin{equation}
M_{RR} = (0.23\pm0.04)({\rm [Fe/H]} + 1.5) + (0.59\pm0.03)\label{abs_mag},
\end{equation}
where ${\rm [Fe/H]}$ is the spectroscopic metallicity of an RR Lyrae star (see
Section~\ref{metallicities}). Following the discussion by \citet{vz06} (see
their Section 4), we adopt
$\sigma_{M_V}=\sqrt{(\sigma_{M_V}^{[Fe/H]})^2+(\Delta M_V^{ev})^2}=0.09$ mag as
the uncertainty in absolute magnitude, where $\sigma_{M_V}^{[Fe/H]}=0.05$ mag is
the uncertainty in Equation~\ref{abs_mag} due to uncertainty in measured
spectroscopic metallicity ($\sigma_{[Fe/H]}=0.15$ dex, see
Section~\ref{metallicity}), and $\Delta M_V^{ev}=0.08$ mag is the uncertainty
due to RR Lyrae evolution off the zero-age horizontal branch \citep{vz06}.
Combined with a maximum of 0.05 mag uncertainty in the flux-averaged magnitude
$\langle m\rangle$, we find the uncertainty in the distance modulus to be
$\sim0.1$ mag, or $\sim5\%$ in distance.

\subsection{Light Curve Parameters of CSS RR Lyrae Stars}

The light curve parameters of PTF and LINEAR RRab stars were simply adopted from
papers describing their selection. However, the same could not be done for CSS
RRab, as we describe below.

Following a visual inspection of phased light curves of CSS RRab stars (phased
using periods and ephemeris values taken from Table~1 of \citealt{dra13}), we
noticed that the phased light curves did not peak at phase of zero, as is the
convention for RR Lyrae stars, but at phases that varied from star to star. This
prevents accurate determination of the phase of pulsation, which is important
when subtracting the velocity due to pulsations from the measured radial
velocity (see Section~\ref{radial_velocities}). We traced this problem to
incorrect ephemeris values listed in Table~1 of \citealt{dra13}, and notified
Drake et al.~of this issue.

To determine the correct epoch of maximum light for CSS RRab stars, we fitted
SDSS $r$-band RRab light curve templates of \citet{ses10} to phased CSS light
curves. The light curve parameters obtained using this procedure, such as the
epoch of maximum light (needed for accurate estimation of the phase) and the
flux-averaged magnitude (needed for distance estimates), are listed in
Table~\ref{table-positions} and are used hereafter instead of the ones provided
by \citet{dra13}. For comparison, the newly derived light curve amplitudes and
the ones listed in Table~1 of \citet{dra13} agree to 3\%.

To support future spectroscopic followup of RRab stars in the vicinity of the
Orphan stream, positions and light curve parameters of CSS RRab stars without
spectroscopic observations (open squares in Figures~\ref{fig1} and~\ref{fig2})
are provided in Table~\ref{css_rr}.

\subsection{Spectroscopic Observations}

The spectroscopic observations were obtained using the blue channel of the
Double Spectrograph (DBSP; \citealt{og82}) mounted on the Palomar 5.1-m
telescope. A 600 lines mm$^{-1}$ grating and a {5600 \AA} dichroic were used,
providing a resolution of $R=1360$ and a spectral range from 3800 {\AA} to
{5700 \AA}.

Target RR Lyrae stars were observed over the course of several nights in 600 s
to 900 s exposures. In order to avoid the discontinuity in the radial velocity
curve near maximum light, the observations were scheduled to target stars
between phases of 0.1 and 0.85 of their pulsation cycle, with earlier phases
being preferred as the stars are brighter then.

In addition, we also targeted several equivalent-width (EW) standard stars (see
Table~6 of \citealt{lay94}), as well as a few bright RR Lyrae stars with
well-determined spectroscopic metallicities. The equivalent-width standard stars
were observed in order to transform EWs measured from DBSP spectra to Layden's
(1994) system (see his Section 3.4.3 and our Section~\ref{metallicities} below).
The bright RR Lyrae stars (VY Ser, RR Lyr, ST Boo, and VX Her) were observed in
order to validate the metallicity measurements described in
Section~\ref{metallicities}.

\subsection{Data Reduction and Calibration}\label{reduction}

All data were reduced with standard IRAF\footnote{IRAF is distributed by the
National Optical Astronomy Observatory, which is operated by the Association of
Universities for Research in Astronomy (AURA) under cooperative agreement with
the National Science Foundation (\url{http://iraf.noao.edu}).} routines, and
spectra were extracted using an optimal (inverse variance-weighted) method
\citep{hor86}. The wavelength calibration of spectra was done using a set of
FeAr arc lines, and the spectra were flux-calibrated. The signal-to-noise ratio
of the spectra ranged from 10 to 30 at 4750 \AA.

To correct for the possible wavelength shift in spectra during an exposure
(e.g., due to instrument flexure), we measured the wavelengths
($\lambda_{\rm obs}$) of three ${\rm[Hg\, I]}$ sky lines and one ${\rm[O\, I]}$
sky line, and then adjusted the zero-point of each spectrum by the mean of
$\Delta\lambda=\lambda_{\rm obs}-\lambda_{\rm lab}$, where $\lambda_{\rm lab}=$
4046.565, 4358.335, 5460.750, and {5577.340 \AA} are the laboratory wavelengths
of ${\rm [Hg\, I]}$ and ${\rm[O\, I]}$ sky lines when observed in air. The
uncertainty in the zero-point of wavelength calibration was estimated as the
standard error of the mean of $\Delta \lambda$ (i.e., standard deviation of
$\Delta \lambda$ divided by two). On average, this uncertainty is $\sim0.08$
{\AA}, or $\sim5$ km s$^{-1}$ at 4750 {\AA}.

\subsection{Center-of-mass Velocities}\label{radial_velocities}

The center-of-mass velocity (hereafter, the systemic velocity $v_{helio}$), is
the line-of-sight velocity of an RR Lyrae star that one would measure if its
atmosphere was at rest. As described below, we first measure the heliocentric
radial velocities (RVs) of ${\rm H\gamma}$ and ${\rm H\beta}$ Balmer lines
($v_{\rm H\gamma}$ and $v_{\rm H\beta}$), and then fit them to template RV
curves as a function of phase to obtain the systemic velocity. We use Balmer
lines because they are least affected by blending in low resolution DBSP
spectra. The velocities are measured separately for each Balmer line because the
lines form at different heights in the atmosphere and have different velocities
as a function of pulsation phase \citep{oke62, sesar12}.

The RVs were measured by cross-correlating observed spectra with about 350
synthetic spectra (hereafter, template spectra) selected from the \citet{mun05}
spectral library\footnote{\url{http://archives.pd.astro.it/2500-10500/}}. The
template spectra have a linear dispersion of 1 {\AA} per pixel and encompass the
range of metallicities, surface gravities and effective temperatures covered by 
RRab stars ($5000 < T_{eff}{/\rm K} < 10000$, $1.5 < \log g < 3.5$,
$-2.5 < {\rm [Fe/H]/dex} < -0.5$).

The cross-correlation was done separately for two spectral regions, one centered
on ${\rm H\gamma}$ (4160--4630 \AA) and the other one centered on ${\rm H\beta}$
(4630--5000 \AA), using an IDL program written by V.~Bhalerao
({\em getvel}\footnote{http://www.iucaa.ernet.in/$\sim$varunb/getvel/};
\citealt{bha12}). Briefly, this program takes a template spectrum, resamples it
to the dispersion of the observed spectrum, convolves the resampled spectrum
with an appropriate Line Spread Function, and finds the velocity shift for which
the $\chi^2$ per degree of freedom between the template and the observed
spectrum is the lowest. The velocity obtained from the template with the lowest
$\chi^2$ per degree of freedom is adopted as the best-fit RV ($v_{\rm H\gamma}$
or $v_{\rm H\beta}$). The uncertainty in the best-fit RV, $\sigma_{cc}$, was
estimated as the velocity range around the best-fit RV within which the
$\chi^2$ per degree of freedom increases by 1. On average, this uncertainty is
about 13 km s$^{-1}$.

The systemic velocities were determined by fitting template RV curves to
measured ${\rm H\gamma}$ and ${\rm H\beta}$ velocities. For this purpose, we
used template ${\rm H\gamma}$ and ${\rm H\beta}$ velocity curves constructed by
\citet{sesar12}. Given an RRab star with a Johnson $V$-band light curve
amplitude of $A_V$, the amplitude of a Balmer line RV template ($A_{rv}$) was
set using relations of \citet{sesar12} 
\begin{eqnarray}
A_{rv}^{H\gamma} = 46.1(\pm2.5)A_V + 38.5(\pm2.4),\, \sigma_{fit}=2.8\, {\rm km\, s^{-1}}\label{Hgamma}\\
A_{rv}^{H\beta} = 42.1(\pm2.5)A_V + 51.1(\pm2.4),\, \sigma_{fit}=3.0\, {\rm km\, s^{-1}}\label{Hbeta}
\end{eqnarray}
The scaled template was then shifted in velocity to match the $v_{\rm H\gamma}$
or $v_{\rm H\beta}$ measurement at the corresponding phase; the systemic
velocity is that of the shifted template at phase 0.5. The $V$-band light curve
amplitudes of observed RRab stars were calculated from their $R$-band light
curve amplitudes as $A_V = 1.21 A_R$ (see Section 5 of \citealt{sesar12}).

For each star, the above procedure returns two estimates of the systemic
velocity, one based on ${\rm H\gamma}$ ($v_{helio,H\gamma}$) and the other one
based on ${\rm H\beta}$ radial velocity ($v_{helio,H\beta}$). Following
\citet{sesar12}, the variance of a systemic velocity was calculated as
\begin{align}
\sigma^2_{v} &= \sigma^2_{cc} + \sigma^2_{model} \\
&= \sigma^2_{cc} + (A_{rv}+\sigma_{fit})^2[\sigma^2_{template}(\Phi_{obs}) + (0.1k)^2] \nonumber,
\end{align}
where $k=1.54$ and $k=1.42$ for ${\rm H\beta}$ and ${\rm H\gamma}$ RV templates,
respectively. The $\sigma^2_{template}(\Phi_{obs})$ term is the variance of the
template at the phase of observation (see bottom plots in Figure~1 of
\citealt{sesar12}), and $\sigma^2_{cc}$ is the variance of the best-fit
heliocentric radial velocity ($v_{\rm H\gamma}$ or $v_{\rm H\beta}$). On
average, $\sigma_{model}$ is about 13 km s$^{-1}$.

The final systemic velocity of an RRab star was obtained by averaging the
estimates based on two Balmer lines, where each estimate was weighted by the
inverse of its variance
\begin{equation}
v_{helio} = \frac{v_{helio,H\gamma}/\sigma^2_{v,H\gamma} + v_{helio,H\beta}/\sigma^2_{v,H\beta}}{1/\sigma^2_{v,H\gamma} + 1/\sigma^2_{v,H\beta}}.
\end{equation}
The uncertainty in the final systemic velocity was calculated
by adding (in quadrature) the uncertainty in the zero-point of wavelength
calibration ($\sigma_{zpt}$) and the uncertainty in the weighted mean (i.e.,
$v_{helio}$)
\begin{equation}
\sigma_{helio} = \sqrt{\sigma^2_{zpt} + \frac{1}{1/\sigma^2_{v,H\gamma} + 1/\sigma^2_{v,H\beta}}}.
\end{equation}
The final systemic velocities and their uncertainties are listed in
Table~\ref{table-results}. In addition to line-of-sight velocities,
Table~\ref{table-results} also lists proper motions taken from the \citet{mun04}
catalog.

\subsection{Spectroscopic Metallicities}\label{metallicities}

Spectroscopic metallicities were measured following the method and calibration
of \citet{lay94} which involves comparing the pseudo-equivalent width of
${\rm [Ca\, II]}$ K line, W(K), against the mean pseudo-equivalent widths of
$\beta$, $\gamma$, and $\delta$ Balmer lines, W(H). 

The pseudo-equivalent widths (hereafter, EWs) of the ${\rm [Ca\, II]}$ K line
and Balmer lines were measured from DBSP spectra (normalized to the
pseudo-continuum) using the EWIMH program\footnote{\url{http://physics.bgsu.edu/~layden/ASTRO/DATA/EXPORT/EWIMH/ewimh.htm}} written by A.~Layden. Measured EWs
({\rm $W^\prime(K)$}, {\rm $W^\prime(H\delta)$}, {\rm $W^\prime(H\gamma)$}, and
{\rm $W^\prime(H\beta)$}) were then transformed to Layden's (1994) EW system
using the following relations:
\begin{align}
W(K) &= 1.11 W^\prime(K) - 0.24\label{K_DBSP}\\
W(H\delta) &= 0.87 W^\prime(H\delta) + 1.02 \\
W(H\gamma) &= 1.26 W^\prime(H\gamma) - 1.00 \\
W(H\beta) &= 1.02 W^\prime(H\beta) + 1.22.
\end{align}
The above equations were derived by comparing EWs of eight equivalent-width
standard stars observed with DBSP, with the EW values measured by \citet{lay94}
and listed in his Table~6. After the transformation, W(K) was corrected for
interstellar ${\rm [Ca\, II]}$ absorption using the \citet{bee90} model
\begin{equation}
W(K_0) = W(K) - W_{max}\left(1-e^{-|z|/h}\right)/\sin|b|,
\end{equation}
where $W_{max}=0.192$ \AA, $h=1.081$ kpc, $b$ is the Galactic latitude, and
$z$ is the height above the Galactic plane in kpc.

The spectroscopic metallicity was calculated as
\begin{equation}
[Fe/H] = \frac{W(K_0) - a - bW(H)}{c+dW(H)}\label{spec_feh},
\end{equation}
where $a = 13.858$, $b = -1.185$, $c = 4.228$, $d = -0.32$ (see Table~8 in
\citealt{lay94}). Equation~\ref{spec_feh} was obtained by inverting Equation 7
of \citet{lay94}. The metallicities measured using this method are listed in
Table~\ref{table-results}.

To validate the measured metallicities, we observed four bright RR Lyrae stars
(VY Ser, RR Lyr, ST Boo, and VX Her) that have well-determined spectroscopic
metallicities ($-1.9 < {\rm [Fe/H]/dex} < -1.3$). An unweighted linear
least-squares fit between metallicities measured by us and by \citet{lay94}
returned a slope consistent with one (1.05), and an intercept consistent with
zero (0.01). The root-mean-square (rms) scatter of the fit was 0.07 dex, which
we adopt as the systematic uncertainty of this method. Thus, we conclude that
our metallicities are on the \citet{lay94} metallicity system, which is tied to
the \citet{zw84} globular cluster abundance scale.

\subsection{Comparison with SDSS Measurements}

SDSS and SEGUE spectra exist for 27 RRab stars from our spectroscopic sample.
For these spectra, the SEGUE Stellar Parameters Pipeline
\citep[SSPP;][]{lee08,lee11} provides metallicities and radial velocities.
Before we compare our measurements to the ones supplied by SSPP, there are three
important points that need to be made.

First, SDSS spectra from which SSPP measurements are derived consist of
multiple exposures, usually taken back-to-back but sometimes even spread over
days \citep{bic12} (i.e., obtained at vastly different pulsation phases). Since
the exposures were taken without the knowledge of the pulsation phase, some of
the exposures may have been acquired during the rapid expansion phase (between
phases of 0.9 and 1.0). The position and the width of spectral lines change
rapidly during this phase and the measurements obtained from such spectra may
not be reliable. Second, SSPP measures radial velocities by fitting a suitable
spectral template to the entire observed (coadded) spectrum. This is not an
optimal approach because the observed velocities measured from metallic and
Balmer lines exhibit different behavior as a function of phase
\citep{oke62, sesar12}. Ideally, the systemic velocity should be determined for
each line separately and then multiple estimates should be averaged out, as done
in our Section~\ref{radial_velocities}. And third, SSPP uses 12 separate methods
to measure ${\rm [Fe/H]}$, none of which matches the one used in
Section~\ref{metallicities}. The SSPP combines ${\rm [Fe/H]}$ values from the
various methods and provides an overall best value (FEHADOP) along with an
uncertainty.

Out of the 27 RRab stars observed by us and by SDSS, only 10 stars have coadded
SDSS spectra where the difference between the start of the first exposure and
the end of the last exposure is less than an hour. The total exposure time of
these coadded spectra is short enough ($<10\%$ of the period) that the blurring 
of spectral lines is not significant. We corrected SSPP radial velocities for
pulsation using corrections for the $H\gamma$ line (these corrections gave the
smallest scatter with respect to systemic velocities determined in
Section~\ref{radial_velocities}). The average difference between our and SSPP
velocities is $\sim-12$ {\kms} and the rms scatter of differences is $\sim20$
{\kms}. This scatter is fully consistent with uncertainties in estimated
systemic velocities. Out of 10 RRab stars with good SDSS spectra, only 7 stars
have ${\rm [Fe/H]}$ measured by SSPP. The average difference between the
${\rm [Fe/H]}$ measured in Section~\ref{metallicities} and by SSPP is -0.36 dex
and the rms scatter is 0.15 dex (i.e., SSSP metallicities are higher and the
scatter is consistent with uncertainties). This value is similar to the -0.4 dex
offset found by \citet[see their Section~6.2.1]{dra13}.

In principle, we could average out our metallicity and velocity measurements
with the ones determined by SSPP. However, adding SSPP measurements would not
change any of the conclusions reported in following Sections (e.g., the
likelihood of membership) and may introduce systematic uncertainties which were
not uncovered in the simple comparison presented in this Section. Thus, we
choose not to use SSPP data at this moment and plan to revisit this issue once
properly measured SSPP velocities and metallicities become available for RRab
stars (N.~De Lee, in preparation).

\section{Results}\label{results}

\begin{figure*}
\plottwo{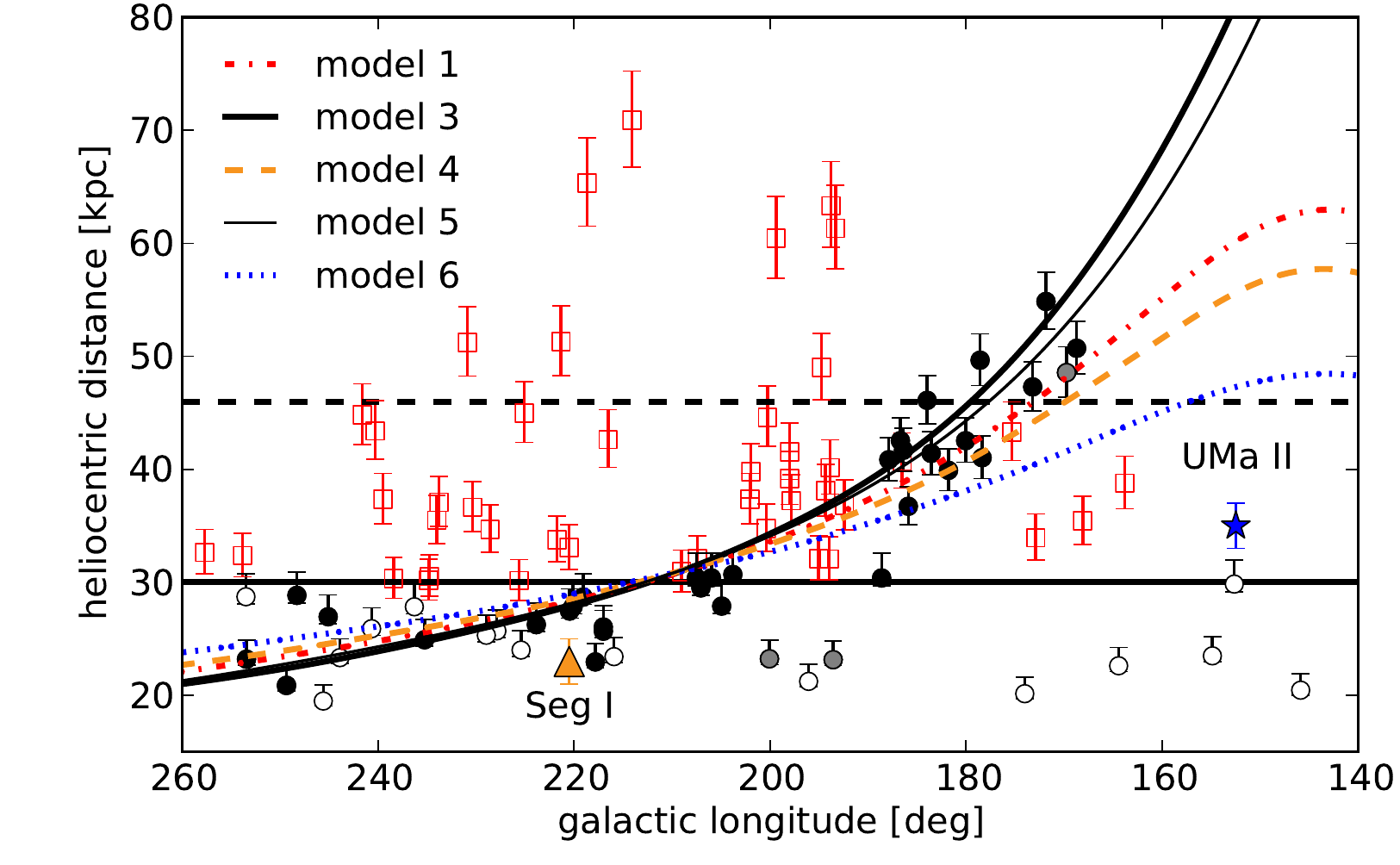}{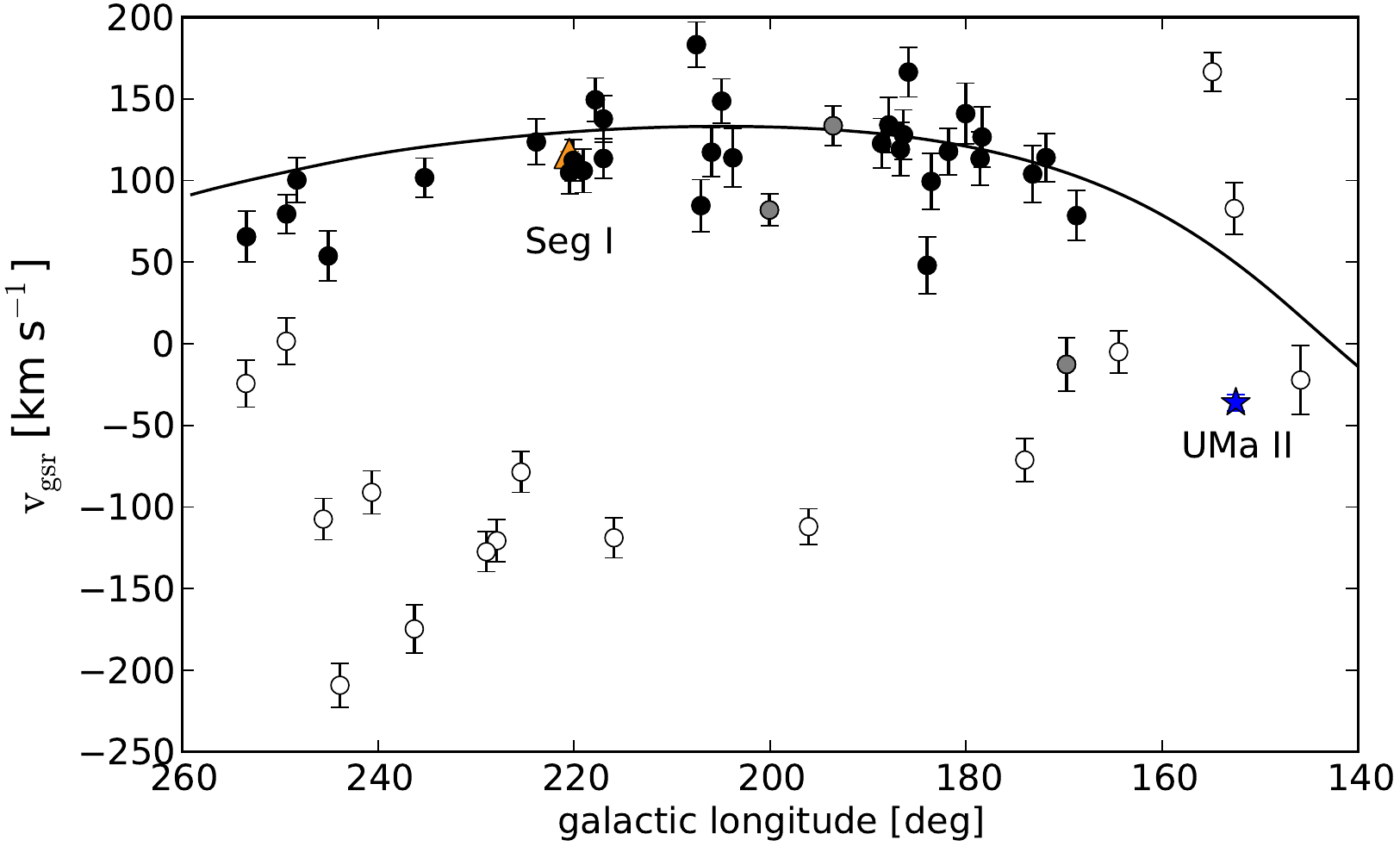}

\plottwo{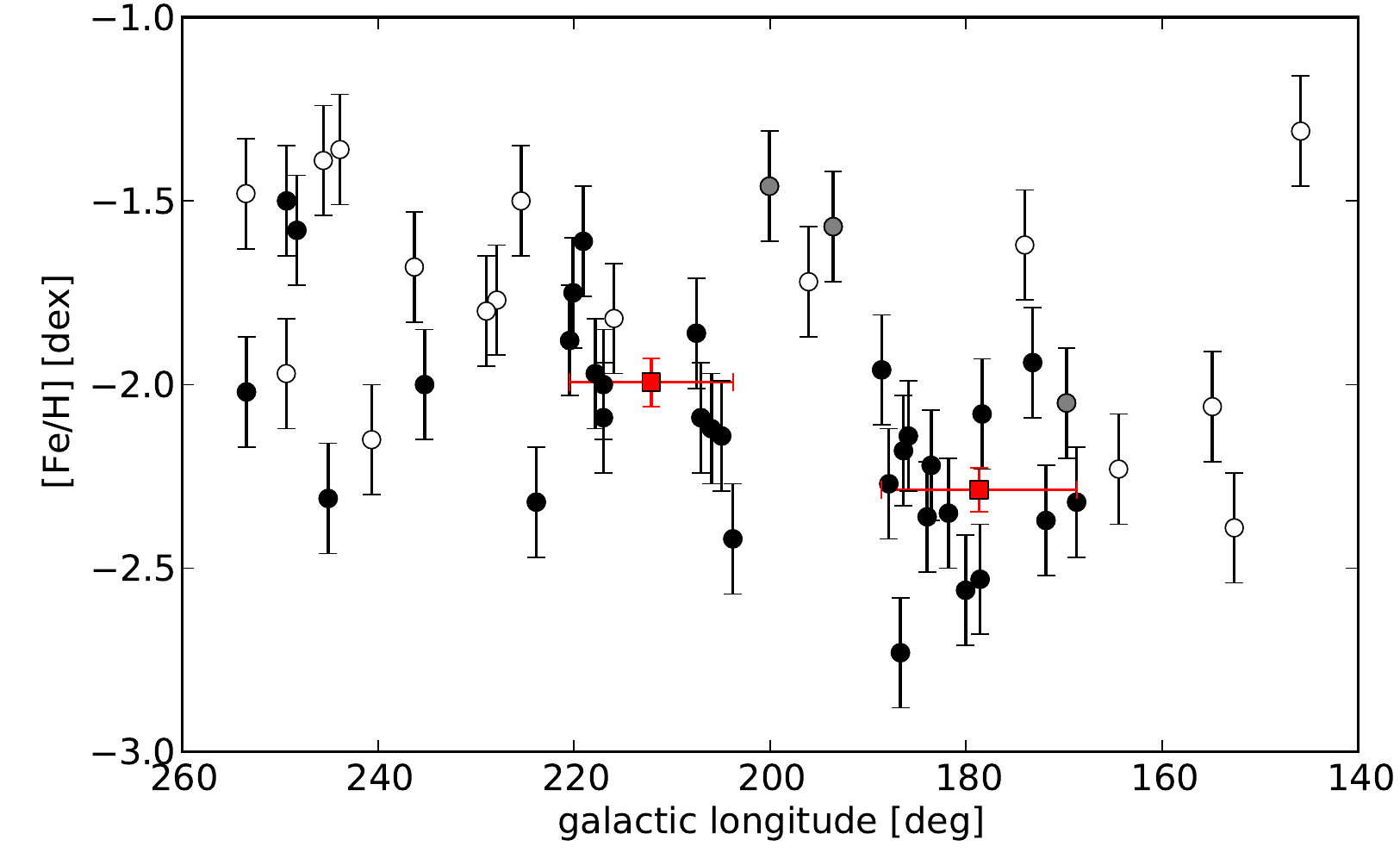}{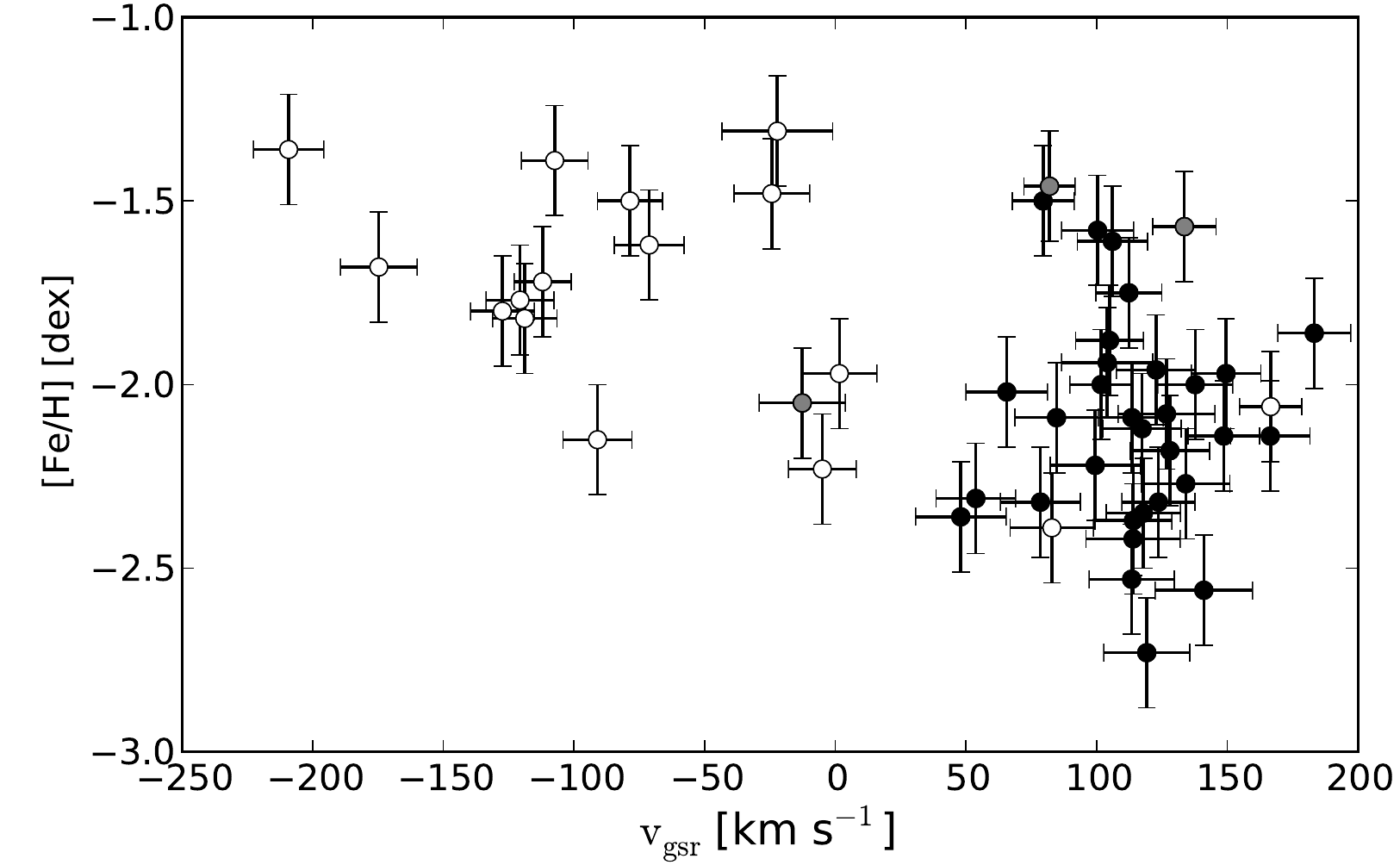}
\caption{
{\em Top left}: Heliocentric distances vs.~galactic longitude, for RRab stars in
the vicinity of the Orphan stream. The meaning of symbols is the same as in
Figure~\ref{fig1}. The error bars reflect the uncertainty of 0.1 mag in distance
modulus, or $\sim5\%$ in distance. The heliocentric distances for CSS RRab stars
without spectroscopic observations ({\em open squares}) were calculated assuming
${\rm [Fe/H]} = -2.0$ dex (uncertainty in distance is $\sim7\%$). The
heliocentric distances of Segue I and Ursa Major II were taken from
\citet{dal12} and \citet{bel07b}, respectively. The lines show
some\footnote{The lines for models 2 and 7 are not shown due to their similarity
to lines for models 3 and 5, respectively.} of the best-fit orbits of the Orphan
stream obtained by \citet{new10}, assuming different models for the potential
(see Table~\ref{table-models} and Section~\ref{comparison} for a description of
their models). The solid horizontal line at 30 kpc shows the distance beyond
which the number density of stars associated with the ``smooth'' stellar halo
rapidly decreases \citep{sji11}, and the dashed horizontal line at 46 kpc shows
the maximum distance probed by \citet{new10}. {\em Top right}: Galactic standard
of rest velocities vs.~galactic longitude. The line shows the best-fit orbit of
\citet{new10} assuming a potential defined by their model 3. For clarity, other
orbits of \citet{new10} are not plotted because they are indistinguishable from
the plotted one given the range of galactic longitudes probed here and given the
uncertainties in velocities of RR Lyrae stars. The majority of stars associated 
with the Orphan stream have $v_{gsr}\sim100$ {\kms}. The stars with
$v_{gsr}\sim-120$ {\kms} are likely associated with the Sagittarius tidal
stream. The velocities of Segue I (orange triangle at $l\sim220\arcdeg$) and
Ursa Major II (blue star symbol) were taken from \citet{mar08} and
\citet{mar11}, respectively. {\em Bottom left}: Metallicity vs.~galactic
longitude. The error bars are set to 0.15 dex. The squares show the mean
metallicity for two groups of RR Lyrae stars with high likelihoods of being
members of the Orphan stream. The vertical error bars show the standard error of
the mean and the horizontal error bars indicate the width of bins. The
difference between the two mean metallicities is 0.3 dex, suggesting a possible
metallicity gradient (the standard error of the mean is $\sim0.06$ dex).
{\em Bottom right}: Metallicity vs.~the velocity in the Galactic standard of
rest. Note how the velocity is a better criterion for separating Orphan and
non-Orphan RR Lyrae stars than the metallicity. RR Lyrae stars associated with
the Orphan stream are on average metal-poor (mean ${\rm [Fe/H]=-2.1}$ dex), and
have metallicities from -1.5 dex to -2.7 dex (standard deviation is 0.3 dex).
The group of stars with $v_{gsr}\sim -120$ {\kms} and ${\rm [Fe/H]=-1.8}$ dex
has the velocity and the metallicity consistent with that of the Sagittarius
leading tidal stream \citep{yan09b}.
\label{fig2}} 
\end{figure*}

Using data gathered and described in Section~\ref{data}, we can now study
heliocentric distances, velocities, and metallicities of RR Lyrae stars in the
vicinity of the Orphan stream. To allow a direct comparison with \citet{new10}
results, we adopt their definition and calculate the velocities in the Galactic
standard of rest ($v_{gsr}$) as
\begin{equation}
v_{gsr} = v_{helio} + 10.1\cos b \cos l + 224\cos b \sin l + 6.7\sin b.
\end{equation} 

Following \citet{new10}, we have plotted the heliocentric distances, velocities,
and metallicities as a function of galactic longitude $l$ in Figure~\ref{fig2}.
In subsequent Sections, we use this figure to first select likely members of the
Orphan stream, and then to examine various other properties of the stream, such 
as its metallicity and spatial extent.

\subsection{Orphan Stream Members}\label{members}

In this Section, we characterize the likelihood that an RRab star is a member of
the Orphan stream. As we describe below, a star is assigned a high likelihood of
being an Orphan stream member if
\begin{enumerate}
\item Its galactic longitude is within $160\arcdeg<l<260\arcdeg$ and its sky
position is within $4\arcdeg$ from the equator of the Orphan stream
($|B_{Orphan}| < 4\arcdeg$), and
\item Its velocity in the Galactic standard of rest is $v_{gsr}>40$ {\kms}.
\end{enumerate}
The heliocentric distance of a star is only a weak criterion and is used to
increase (from low) or decrease (from high) the probability of three RRab stars
to medium probability (see below). We emphasize that only RRab stars with high
likelihood of membership are used when analyzing the metallicity of the stream
(Section~\ref{metallicity}), its extent (Section~\ref{extent}), and when making
comparisons with models (Section~\ref{comparison}).

Looking at the top left panel of Figure~\ref{fig2}, we can see that the best-fit
orbits predict an increase in the heliocentric distance of the stream as the
galactic longitude decreases, with the stream moving above 30 kpc for
$l<210\arcdeg$. At distances beyond 30 kpc, the number density of halo stars
associated with the ``smooth'' stellar halo rapidly decreases
\citep{ses10, sji11, dbe11}, and as a result, halo substructures such as stellar
streams are easier to discern (e.g., see Figure~11 of \citealt{ses10}). Indeed,
RR Lyrae stars in the vicinity of the Orphan stream and with $l<210\arcdeg$ seem
to be distributed in an arc-like structure, which would be discernible even if
we had no knowledge of the Orphan stream. The majority of stars in this region
also have positive velocities in the Galactic standard of rest, as evident in
the top right panel of Figure~\ref{fig2} ($v_{gsr}>40$ {\kms}), and these
velocities agree with values observed by \citet{new10}. Thus, we tag these stars
as having a high probability of being Orphan stream members (see the last column
in Tables~\ref{table-positions} and~\ref{table-results}).

For $l\lesssim210\arcdeg$, and according to predicted orbits, the stream's
heliocentric distance is expected to dip below 30 kpc. At these distances, the
number density of RR Lyrae stars associated with the ``smooth'' halo is expected
to be non-negligible, making the association of stars with the Orphan stream
less straightforward. In this region, we rely on velocities to assign
probabilities, and tag stars with $v_{gsr}>40$ {\kms} as having a high
probability of being Orphan stream members.

For the most part, the stars can be easily tagged as having a high or low
membership probability based on their position and velocity. However, there are
three stars that were tagged as having a medium probability. RR5 is located at
$l\sim170\arcdeg$ and $D\sim48$ kpc, but has a very low velocity compared to
other stars at this position ($v_{gsr}\sim-2$ {\kms} vs.~$v_{gsr}\sim100$
{\kms}). Since RR Lyrae stars are standard candles, the heliocentric distance is
considered reliable; this star is likely located at $D>40$ kpc even if we assume
its metallicity to be ${\rm [Fe/H]}= -1.0$ dex. In the case of this star, the
measured velocity is more likely to be incorrect than the distance, and
additional velocity observations may resolve this discrepancy. If the currently
measured velocity is wrong, then we believe this star is a likely member of the 
Orphan stream. 

Unlike RR5, RR20 and RR22 have velocities consistent with those of Orphan stream
stars at $l\sim200\arcdeg$ ($v_{gsr}\sim100$ {\kms}), but are much closer than
the predicted orbit of the stream (23 kpc vs.~35 kpc for the stream). In
addition, these stars have a metallicity of ${\rm [Fe/H]\sim-1.6}$ dex, making
them rather metal-rich for the Orphan stream. Unless the Orphan stream has a
much greater line-of-sight breadth than previously thought, we consider these
stars as less likely members of the Orphan stream. As for CSS RRab stars without
spectroscopic observations, shown as open squares in Figure~\ref{fig1} and the
top left panel of Figure~\ref{fig2}, we cannot estimate likelihood of their
membership without spectroscopic data (i.e., radial velocities), even for stars
that are seemingly near the predicted orbit of the Orphan stream. Stars without
spectroscopic observations are not used in our analysis and their followup is
encouraged.

We also considered using proper motions as additional criteria for assigning
membership probabilities. Unfortunately, the range of proper motions for likely 
and unlikely members was too small relative to the uncertainty in measurements
to make proper motions useful for this purpose ($|\mu_{l,b}|\lesssim5$ mas
yr$^{-1}$ vs.~$|\mu_{err}|\sim3$ mas yr$^{-1}$).

\subsection{Metallicity of the Orphan Stream}\label{metallicity}

Since the metallicity was not used as a criterion while assigning membership
probabilities, it would be interesting to examine the distribution of
metallicities for stars that are likely or unlikely to be associated with the
Orphan stream.

The bottom left panel of Figure~\ref{fig2} shows the metallicity of RR Lyrae
stars along the Orphan stream (i.e., in the direction of galactic longitude).
Here we have assumed that the uncertainty in metallicity is 0.15 dex for all
stars. We consider this error to be slightly overestimated. As stated in
Section~\ref{metallicities}, a comparison of measured and literature values of
metallicities of four bright RR Lyrae stars indicates that the systematic
uncertainty in metallicity is about 0.07 dex. Furthermore, one of the stars in
our sample (RR6) was observed on two different nights\footnote{Its velocity and
the metallicity listed in Table~\ref{table-results} are average values.}, and
the measured metallicities were -2.45 dex and -2.29 dex. The rms scatter of
these two measurements is 0.11 dex, even though the SNR per {\AA} of both
observations was only 12. 

As the bottom left panel shows, likely members of the Orphan stream span a wide
range of metallicities, from -1.5 dex to -2.7 dex. Stars that are unlikely to
be associated with the Orphan stream also span a wide range, from -1.3 dex to
-2.3 dex. The two samples overlap significantly in metallicity, making the
metallicity a weak criterion for identifying members of the Orphan stream. The
velocity in the Galactic standard of rest is a more powerful criterion, as
evident from the bottom right panel of Figure~\ref{fig2}.

The metallicity vs.~position panel does show an interesting trend. While the
stars not associated with the Orphan stream seem to span a wide range of
metallicities regardless of their position, stars associated with the stream
seem to be increasingly more metal-rich in the eastward direction. To verify
this, we have binned likely members into two groups and calculated the mean
metallicity and the standard error of the mean for each group. We find that the
stars located near $l\sim210\arcdeg$ are 0.3 dex more metal-rich than the stars
near $l\sim180\arcdeg$ (the standard error of the mean is $\sim0.06$ dex).

The statistical dependence between the galactic longitude of a star and its
metallicity can be tested using the Kendall's $\tau_B$ coefficient (e.g., see
\citealt{lup93}). This test is non-parametric, as it does not rely on any
assumptions on the distributions of X or Y or the distribution of (X,Y). Values
of $\tau_B$ range from -1 (perfect inversion) to 1 (perfect agreement). A value 
of zero indicates the absence of association. Using 27 highly likely members
with $l<240\arcdeg$ (to minimize possible contamination by stars associated with
the smooth halo spheroid, see below), we obtained $\tau_B=0.416\pm0.002$,
indicating statistically significant correlation between galactic longitude of
an Orphan stream RRab star and its metallicity. The correlation between
${\rm [Fe/H]}$ and $\Lambda_{Orphan}$ (position along the Orphan stream) is
slightly stronger and is $\tau_B=0.433\pm0.002$.

To estimate the probability that the observed correlation is due to random
chance, we generate 10,000 samples of 27 stars and count the number of samples
where $\tau_B>0.416$ and the uncertainty in $\tau_B < 0.002$. The samples are
generated by drawing metallicities from a 0.3 dex-wide Gaussian centered on -2.1
dex (i.e., the metallicity distribution of highly likely Orphan stream members),
while keeping the galactic longitude of stars unchanged. We find that the
probability of seeing the observed correlation due to random chance is
$\lesssim10^{-3}$. The same result is obtained if instead of drawing
metallicities, we randomly exchange positions of stars used in this analysis.

It is not impossible that the observed ${\rm [Fe/H]}$ vs.~$l$ correlation is due
to unidentified contamination by RRab stars actually associated with the smooth
halo spheroid. Since the metallicity of halo stars is centered on
${\rm [Fe/H]}=-1.5$ dex \citep{ive07}, unidentified contamination could make the
Orphan stream seem more metal-rich as it gets closer to the Galactic plane, that
is, as the galactic longitude increases. Because the unidentified contamination
and a metallicity gradient could both make the Orphan stream seem more
metal-rich, we cannot use the ${\rm [Fe/H]}$ vs.~$l$ distribution (i.e., the
bottom left panel of Figure~\ref{fig2}) to make any estimates of the
contamination. Velocities, however, should not be affected by the metallicity
gradient, which is why they can be used to make statements on the level of
contamination.

Based on the velocity distribution of stars in the top right panel of
Figure~\ref{fig2}, we do not think the metallicity gradient is caused by
unindentified contamination by halo stars. If the contamination was significant,
then there would be more RRab stars with $v_{gsr}\sim0$ {\kms}, since the radial
velocity distribution of halo stars is a $\sim100$ {\kms}-wide Gaussian centered
on zero  \citep{bon10}. Instead, the distribution of velocities in the top right
panel of Figure~\ref{fig2} is clearly bimodal, with stars having velocities at
the extreme of the halo velocity distribution. The only exception may be stars
at $l>240\arcdeg$, which is why we excluded them when studying the correlation
between the galactic longitude of an Orphan stream RRab star and its
metallicity.

In Section~\ref{metallicities} we stated that the measured ${\rm [Ca\, II]}$
pseudo-equivalent widths, W(K), were corrected for interstellar
${\rm [Ca\, II]}$ absorption using the \citet{bee90} model. At large distances
from the Galactic plane, this model is equal to $W(K_{ism})=0.192/\sin(|b|)$,
where $b$ is the galactic latitude. Since the two groups of stars are at
different galactic latitudes ($48\arcdeg$ and $53\arcdeg$), it is not impossible
that the observed gradient is caused by this correction. To verify whether this
is the case, we estimated the metallicities without correcting the W(K) for
interstellar ${\rm [Ca\, II]}$ absorption, and compared them to values listed in
Table~\ref{table-results}. The metallicities that were not corrected for
interstellar ${\rm [Ca\, II]}$ absorption were higher by 0.1 dex, but that
offset was the same for both groups. Thus, we conclude that the observed
gradient is not due to the correction for interstellar ${\rm [Ca\, II]}$
absorption.

\subsection{Does the Orphan Stream End at 55 kpc?}\label{extent}

One of the goals of this work was to trace the Orphan stream beyond the 46 kpc
limit reached by \citet{new10}. As the top left panel of Figure~\ref{fig2}
shows, we have achieved that goal. The question that we now ask is, does the
Orphan stream extend further into the halo or not?

The question of how far the Orphan stream reaches into the halo is an important
one. As the curved lines in the top left panel of Figure~\ref{fig2} show, the
heliocentric distance of the Orphan stream constrains the halo potential at
large distances. If there are Orphan stream RRab stars beyond 55 kpc, they
should be followed-up spectroscopically because their properties could be used
to even better constrain the halo potential and the orbit of the stream.
Alternatively, a lack of Orphan stream RRab stars beyond 55 kpc may indicate
that we have detected the end of the leading\footnote{Based on the surface
density profile and the velocities of Orphan stream stars, \citet{new10} have
concluded that the part stretching from $l = 250\arcdeg$ to $l=170\arcdeg$ is
the leading tidal arm.} arm of the stream, and such a finding would be important
for future theoretical and observational studies of the stream.

To make sure we did not miss RRab stars due to an overly restrictive spatial
cut, we defined a region on the sky (dotted box in Figure~\ref{fig1}), and
carefully searched the PTF data set in this region for distant RR Lyrae stars
that may have been missed. The result of this search: no additional RRab stars
were identified.

To estimate the incompleteness of our search, we ran our selection algorithm on
a sample of 100 simulated RR Lyrae light curves. The light curves were created
by sampling 26 data points from a template RRab light curve with a maximum
brightness of $m_0 = 18.5$ mag, amplitude of 1.0 mag, and a period of 0.53 days.
These parameters correspond to best-fit parameters of the most distant RRab star
observed in this work (RR6, see Table~\ref{table-positions}). The epoch of
maximum light was randomly selected for each simulated light curve, and the
light curves were sampled at Julian dates of PTF observations in the region of
interest. To make the light curves realistic, we added Gaussian noise to sampled
magnitudes using a model that describes the mean photometric uncertainty in the
PTF $R$-band as a function of magnitude.

We found that an RRab star at $\sim50$ kpc and with an amplitude of 1 mag may
not be recovered 13\% of times (i.e., the recovery fraction is 87\%). A star at 
the same distance, but with a lower amplitude of 0.5 mag may be missed 34\% of
times. At 60 kpc, the incompleteness is 20\% and 46\% for RRab stars with
amplitudes of 1 mag and 0.5 mag, respectively. Thus, it is possible that some
distant RRab stars were not detected due to insufficient signal-to-noise ratio
in PTF data. We intend to obtain additional observations of this region with PTF
to fully explore this possibility.

The boxed region, shown in Figure~\ref{fig1}, does contain a single CSS RRab
star (RR4), which we observed spectroscopically. However, we did not
additionally search CSS in this region for distant RRab stars that may have been
missed by \citet{dra13} because CSS is shallower than PTF. As shown by
\citet{dra13}, their catalog of RR Lyrae stars selected from CSS is about 60\%
complete up to $V\sim17.5$ mag ($\sim25$ kpc), after which completeness
decreases (see their Figure~13). The Mount Lemmon Survey (MLS), which is deeper 
than CSS by $\sim2$ mag \citep{lar07} and which was used by \citet{dra13b} to
discover a potential new stream in the outer halo, does not cover the far end of
the Orphan stream (see Figure 14 of \citealt{dra13b}).

Even if there are no RR Lyrae stars associated with the Orphan stream at
heliocentric distances greater than $\sim55$ kpc, this does not necessarily mean
that the stream ends at $\sim55$ kpc. As the bottom left panel of
Figure~\ref{fig2} shows, the average metallicity of RRab stars in the Orphan
stream seems to be decreasing with decreasing galactic longitude, or
equivalently, with increasing heliocentric distance. At some point, the
metallicity of Orphan stream stars may be low enough that the horizontal branch
of such metal-poor stars is completely devoid of RR Lyrae stars, and is instead 
populated by blue horizontal branch (BHB) stars. BHB stars do not pulsate and we
cannot detect them using our data. Thus, it is possible that the Orphan stream
extends further into the halo, but we are simply unable to trace it there using
RR Lyrae stars.

\subsection{Comparison with Best-fit Orbits of \citet{new10}}\label{comparison}

The positions and velocities of RR Lyrae stars associated with the Orphan stream
allow us to make comparisons with orbits obtained by \citet{new10}. To fit these
orbits, \citet{new10} used seven models for the potential: three models
published by previous authors (models 1, 4, and 6), and four models in which the
halo mass was varied to best match their Orphan stream data (logarithmic and NFW
halos were compared, as well as a low-mass exponential disk versus a high-mass
Miyamoto-Nagai (M-N) disk; see Section 10.1 of \citealt{new10}). Their models
are summarized in Table~\ref{table-models}.

A comparison of heliocentric distances of RRab stars and best-fit orbits is
shown in the top left panel of Figure~\ref{fig2}. The reduced $\chi^2$ values
listed in Table~\ref{table-models} quantify how well the orbits fit distances
of likely RRab members located beyond 33 kpc from the Sun.

The poorest fit is provided by the orbit based on model 6 (reduced $\chi^2=10$).
Model 6 is very similar to model 1, which is based on parameters of
\citet{xue08}, but instead of the $M_{disk}=5\times10^{10}$ M$_\odot$
exponential disk used in model 1, it uses a more massive
$M_{disk}=10\times10^{10}$ M$_\odot$ Miyamoto-Nagai disk. The orbit based on
model 6 clearly does not fit distances of RRab stars located beyond 46 kpc.
Thus, we conclude that the potential of the Milky Way is likely not similar to
the one described by model 6.

The orbits based on models 1, 2, and 3 provide reasonable fits to distances of
remote Orphan stream RRab stars (reduced $\chi^2\sim3$). However, as shown in
Figure~13 of \citet{new10}, these models predict circular velocities near the
Sun's radius that are too low ($V_c < 190$ {\kms}) compared to measurements
obtained by \citet[$V_c=221\pm18$ {\kms}]{krh10} and
\citet[$V_c=218\pm6$ {\kms}]{bov12}.

On the other hand, models 4, 5, and 7 fit the local circular velocity
measurement of \citet{krh10} (see Figure~15 of \citealt{new10}). The distances
of remote Orphan stream RRab stars are also best fit by models 5 and 7 (reduced
$\chi^2=2.3$). Based on this comparison, we conclude that models 5 and 7
represent the best description of the Milky Way potential within 60 kpc (we
have a slight preference for model 7 because it better fits the \citet{krh10}
measurement).

Models 5 and 7 have different halos (logarithmic vs.~NFW), but with the data at
hand and based on best-fit orbits of \citet{new10}, we are not able to
distinguish between these two types of halos. The important characteristic of
these two models is the total mass within 60 kpc, which is
$M_{60}\sim2.7\times10^{11}$ M$_\odot$, or about 60\% of the mass within 60 kpc
found by previous studies \citep{law05,xue08,dea12}.

Compared to heliocentric distances, the velocities of RRab stars (top right
panel of Figure~\ref{fig2}) are not as useful for discriminating models. Near
$l=250\arcdeg$, the median velocity of RRab stars is about 30 {\kms} lower than 
the predicted velocity (75 {\kms} vs.~105 {\kms}). While this is interesting,
note that the sample of Orphan stream RRab stars is quite sparse near this
longitude, so it is difficult to make any strong conclusions.

\section{Discussion and Conclusions}\label{conclusions}

In this work, we used RR Lyrae stars to trace the Orphan stream in the northern
Galactic hemisphere, with a focus on tracing it beyond the 46 kpc limit reached
by \citet{new10}. The main goal of this work was to support future studies of
the Galactic potential by providing a clean sample of Orphan stream RR Lyrae
stars with precise distances and radial velocities (better than 5\% and 15 km
s$^{-1}$, respectively). The secondary goal of this work was to use the
kinematics and positions of RR Lyrae stars associated with the Orphan stream to
see whether new measurements exclude any of the models previously considered by
Newberg et al.

In total, we have observed 50 $ab$-type RR Lyrae stars in the vicinity of the
Orphan stellar stream, and measured their distances, velocities, and
metallicities. Using these data we have classified about 30 RRab stars as being
likely members of the Orphan stream, the most distant of which is located at
about 55 kpc, or about 9 kpc beyond the limit reached by \citet{new10}.

We find that RRab stars in the Orphan stream have a wide range of metallicities,
from -1.5 dex to -2.7 dex. The average metallicity of RRab stars in the Orphan
stream is -2.1 dex (standard deviation is 0.3 dex), and is identical to the
metallicity of BHB stars measured by \citet{new10}. The most distant parts of
the stream (40 to 50 kpc from the Sun) are about 0.3 dex more metal-poor than
the closer parts (within $\sim30$ kpc), suggesting a possible metallicity
gradient along the stream's length. According to the Kendall's $\tau$ test,
the correlation between galactic longitude of an Orphan stream RRab star and its
metallicity seems to be statistically significant ($\tau_B=0.416\pm0.002$).
Using Monte Carlo simulations, we have determined that the probability of seeing
the observed correlation due to random chance is $\lesssim10^{-3}$.

If real, this gradient may reconcile the average metallicity found by
\citet{cas13} (-1.6 dex at Orphan longitude $\Lambda_{Orphan}\sim20\arcdeg$)
with the metallicities measured in this work and by \citet{new10} (-2.1 dex for
$\Lambda_{Orphan}<-10\arcdeg$). Since the only other stellar stream with a known
metallicity gradient is the Sagittarius stream \citep{cho07}, adding the Orphan
stream to this list may provide additional constraints on models of the chemical
evolution of dwarf spheroidal galaxies, evolution of dwarf galaxies, and the
process of tidal stripping and disruption.

The presence of a metallicity gradient could be tested by observing K giants
associated with the Orphan stream. As shown by \citet{cas13}, the most difficult
part of such a study is the identification of likely K giants in the stream. One
way to identify likely candidates would be to use a gravity-sensitive filter
(such as DDO51) in combination with a wide-field imager to separate dwarfs and
giants (e.g., as done by \citealt{mor01}). Once candidate giants are identified,
their classification could be further improved by comparing their position in a
color-magnitude diagram to an isochrone placed at the distance of RRab stars in
the Orphan stream.

Since the heliocentric distance of the Orphan stream can constrain the potential
of the Milky Way (see below), we attempted to trace the stream as far into the
halo as possible. While we successfully mapped the stream to about 55 kpc from
the Sun (about 9 kpc further than \citealt{new10}), we did not identify more
distant RRab stars that could plausibly be Orphan stream members. One
explanation for this may be the incompleteness of our selection, which ranges
from 20\% to 46\% at 60 kpc and for stars with amplitudes from 1 mag to 0.5 mag,
respectively. We intend to obtain additional observations of this region with
PTF to fully explore this possibility. The metallicity gradient could also
explain the lack of distant RRab stars in the Orphan stream. At some point, the 
metallicity of Orphan stream stars may be low enough that the horizontal branch 
of such metal-poor stars is completely devoid of RR Lyrae stars, and is instead
populated by blue horizontal branch (BHB) stars. And finally, the lack of RR
Lyrae stars may indicate that we have detected the end of the leading arm of the
Orphan stream. Detections and follow-up observations of K giants may provide
additional support for this hypothesis.

We have compared the distances of Orphan stream RRab stars with several best-fit
orbits obtained by \citet{new10}, and found that the model 6 of \citet{new10}
cannot explain the heliocentric distances of the most remote Orphan stream RRab
stars. Following similar arguments as \citet{new10}, we find that the distances
of RRab stars prefer potentials where the total mass of the Galaxy within 60 kpc
is $M_{60}\sim2.7\times10^{11}$ $M_\odot$, or about 60\% of that found by
\citet{law05}, \citet{xue08}, and \citet{dea12}. A similar conclusion was also
reached by \citet{new10}.

There is an issue one should be aware of when considering the total mass of the 
Galaxy deduced here. In Section~\ref{introduction}, we said that the
misalignment between the stream and the orbit may be small enough that fitting
an orbit directly to the stream has no significant consequences for the inferred
shape of the Galactic potential \citep{sb13}. However, there may be an
additional misalignment between the stream and the orbit due to the spread in
actions of the stream, caused by a substantial mass and size of the progenitor.
This misalignment may render orbit-fitting, and the potentials inferred from it,
less reliable. In case of the Orphan stream, there seems to be some evidence of
a spread in actions, since the velocity dispersion of BHB stars in the Orphan
stream ($\sim10$ {\kms}, \citealt{new10}) cannot be fully accounted for by the
uncertainty in their radial velocities ($\sim5$ {\kms}). Fortunately,
alternatives to orbit-fitting already exist (e.g., \citealt{sb13b}), and may
yield more reliable results when applied to the Orphan stream.

A better differentiation between various models of the Galactic potential may be
achieved by improving the precision of RRab distances. This can be accomplished 
by observing RRab stars in the near-infrared, where the slope of their
period-luminosity relation (i.e., the Leavitt law) is steeper and width/scatter 
in the relation is smaller \citep{mad12}. The IRAC camera on board the Spitzer
space telescope would be the perfect instrument for this purpose, and we intend 
to propose such a study in the near future.

\acknowledgments

B.S and J.G.C thank NSF grant AST-0908139 to J.G.C for partial support, as do
S.R.K (to NSF grant AST-1009987), and C.J.G (for a NASA grant). E.O.O is
incumbent of the Arye Dissentshik career development chair and is grateful to
support by a grant from the Israeli Ministry of Science and the I-CORE Program
of the Planning and Budgeting Committee and The Israel Science Foundation (grant
No 1829/12). We thank the staff at the Palomar Hale telescope for help and
support with observations. We thank the anonymous referee for the careful
reading of our manuscript and the valuable comments.

This article is based on observations obtained with the Samuel Oschin Telescope
as part of the Palomar Transient Factory project, a scientific collaboration
between the California Institute of Technology, Columbia University, Las Cumbres
Observatory, the Lawrence Berkeley National Laboratory, the National Energy
Research Scientific Computing Center, the University of Oxford, and the Weizmann
Institute of Science. It is also partially based on observations obtained as
part of the Intermediate Palomar Transient Factory project, a scientific
collaboration among the California Institute of Technology, Los Alamos National 
Laboratory, the University of Wisconsin, Millwakee, the Oskar Klein Center, the
Weizmann Institute of Science, the TANGO Program of the University System of
Taiwan, the Kavli Institute for the Physics and Mathematics of the Universe, and
the Inter-University Centre for Astronomy and Astrophysics.

\bibliographystyle{apj}
\bibliography{ms}

\begin{thebibliography}{65}
\expandafter\ifx\csname natexlab\endcsname\relax\def\natexlab#1{#1}\fi

\bibitem[{{Beers}(1990)}]{bee90}
{Beers}, T.~C. 1990, \aj, 99, 323

\bibitem[{{Bell} {et~al.}(2010){Bell}, {Xue}, {Rix}, {Ruhland}, \&
  {Hogg}}]{bel10}
{Bell}, E.~F., {Xue}, X.~X., {Rix}, H.-W., {Ruhland}, C., \& {Hogg}, D.~W.
  2010, \aj, 140, 1850

\bibitem[{{Belokurov} {et~al.}(2007{\natexlab{a}}){Belokurov}, {Evans},
  {Irwin}, {Lynden-Bell}, {Yanny}, {Vidrih}, {Gilmore}, {Seabroke}, {Zucker},
  {Wilkinson}, {Hewett}, {Bramich}, {Fellhauer}, {Newberg}, {Wyse}, {Beers},
  {Bell}, {Barentine}, {Brinkmann}, {Cole}, {Pan}, \& {York}}]{bel07a}
{Belokurov}, V., {Evans}, N.~W., {Irwin}, M.~J., {et~al.} 2007{\natexlab{a}},
  \apj, 658, 337

\bibitem[{{Belokurov} {et~al.}(2007{\natexlab{b}}){Belokurov}, {Zucker},
  {Evans}, {Kleyna}, {Koposov}, {Hodgkin}, {Irwin}, {Gilmore}, {Wilkinson},
  {Fellhauer}, {Bramich}, {Hewett}, {Vidrih}, {De Jong}, {Smith}, {Rix},
  {Bell}, {Wyse}, {Newberg}, {Mayeur}, {Yanny}, {Rockosi}, {Gnedin},
  {Schneider}, {Beers}, {Barentine}, {Brewington}, {Brinkmann}, {Harvanek},
  {Kleinman}, {Krzesinski}, {Long}, {Nitta}, \& {Snedden}}]{bel07b}
{Belokurov}, V., {Zucker}, D.~B., {Evans}, N.~W., {et~al.} 2007{\natexlab{b}},
  \apj, 654, 897

\bibitem[{{Bhalerao}(2012)}]{bha12}
{Bhalerao}, V. 2012, PhD thesis, Caltech

\bibitem[{{Bickerton} {et~al.}(2012){Bickerton}, {Badenes}, {Hettinger},
  {Beers}, \& {Huang}}]{bic12}
{Bickerton}, S., {Badenes}, C., {Hettinger}, T., {Beers}, T., \& {Huang}, S.
  2012, in IAU Symposium, Vol. 285, IAU Symposium, ed. E.~{Griffin},
  R.~{Hanisch}, \& R.~{Seaman}, 289--290

\bibitem[{{Bond} {et~al.}(2010){Bond}, {Ivezi{\'c}}, {Sesar}, {Juri{\'c}},
  {Munn}, {Kowalski}, {Loebman}, {Ro{\v s}kar}, {Beers}, {Dalcanton},
  {Rockosi}, {Yanny}, {Newberg}, {Allende Prieto}, {Wilhelm}, {Lee},
  {Sivarani}, {Majewski}, {Norris}, {Bailer-Jones}, {Re Fiorentin}, {Schlegel},
  {Uomoto}, {Lupton}, {Knapp}, {Gunn}, {Covey}, {Allyn Smith}, {Miknaitis},
  {Doi}, {Tanaka}, {Fukugita}, {Kent}, {Finkbeiner}, {Quinn}, {Hawley},
  {Anderson}, {Kiuchi}, {Chen}, {Bushong}, {Sohi}, {Haggard}, {Kimball},
  {McGurk}, {Barentine}, {Brewington}, {Harvanek}, {Kleinman}, {Krzesinski},
  {Long}, {Nitta}, {Snedden}, {Lee}, {Pier}, {Harris}, {Brinkmann}, \&
  {Schneider}}]{bon10}
{Bond}, N.~A., {Ivezi{\'c}}, {\v Z}., {Sesar}, B., {et~al.} 2010, \apj, 716, 1

\bibitem[{{Bovy} {et~al.}(2012){Bovy}, {Allende Prieto}, {Beers}, {Bizyaev},
  {da Costa}, {Cunha}, {Ebelke}, {Eisenstein}, {Frinchaboy}, {Garc{\'{\i}}a
  P{\'e}rez}, {Girardi}, {Hearty}, {Hogg}, {Holtzman}, {Maia}, {Majewski},
  {Malanushenko}, {Malanushenko}, {M{\'e}sz{\'a}ros}, {Nidever}, {O'Connell},
  {O'Donnell}, {Oravetz}, {Pan}, {Rocha-Pinto}, {Schiavon}, {Schneider},
  {Schultheis}, {Skrutskie}, {Smith}, {Weinberg}, {Wilson}, \&
  {Zasowski}}]{bov12}
{Bovy}, J., {Allende Prieto}, C., {Beers}, T.~C., {et~al.} 2012, \apj, 759, 131

\bibitem[{{Cacciari} \& {Clementini}(2003)}]{cc03}
{Cacciari}, C., \& {Clementini}, G. 2003, in Lecture Notes in Physics, Berlin
  Springer Verlag, Vol. 635, Stellar Candles for the Extragalactic Distance
  Scale, ed. D.~{Alloin} \& W.~{Gieren}, 105--122

\bibitem[{{Casey} {et~al.}(2013){Casey}, {Da Costa}, {Keller}, \&
  {Maunder}}]{cas13}
{Casey}, A.~R., {Da Costa}, G., {Keller}, S.~C., \& {Maunder}, E. 2013, \apj,
  764, 39

\bibitem[{{Chaboyer}(1999)}]{chaboyer99}
{Chaboyer}, B. 1999, Post-Hipparcos Cosmic Candles, 237, 111

\bibitem[{{Chou} {et~al.}(2007){Chou}, {Majewski}, {Cunha}, {Smith},
  {Patterson}, {Mart{\'{\i}}nez-Delgado}, {Law}, {Crane}, {Mu{\~n}oz}, {Garcia
  L{\'o}pez}, {Geisler}, \& {Skrutskie}}]{cho07}
{Chou}, M.-Y., {Majewski}, S.~R., {Cunha}, K., {et~al.} 2007, \apj, 670, 346

\bibitem[{{Dall'Ora} {et~al.}(2012){Dall'Ora}, {Kinemuchi}, {Ripepi},
  {Rodgers}, {Clementini}, {Di Fabrizio}, {Smith}, {Marconi}, {Musella},
  {Greco}, {Kuehn}, {Catelan}, {Pritzl}, \& {Beers}}]{dal12}
{Dall'Ora}, M., {Kinemuchi}, K., {Ripepi}, V., {et~al.} 2012, \apj, 752, 42

\bibitem[{{Deason} {et~al.}(2011){Deason}, {Belokurov}, \& {Evans}}]{dbe11}
{Deason}, A.~J., {Belokurov}, V., \& {Evans}, N.~W. 2011, \mnras, 416, 2903

\bibitem[{{Deason} {et~al.}(2012){Deason}, {Belokurov}, {Evans}, \&
  {An}}]{dea12}
{Deason}, A.~J., {Belokurov}, V., {Evans}, N.~W., \& {An}, J. 2012, \mnras,
  424, L44

\bibitem[{{Drake} {et~al.}(2009){Drake}, {Djorgovski}, {Mahabal}, {Beshore},
  {Larson}, {Graham}, {Williams}, {Christensen}, {Catelan}, {Boattini},
  {Gibbs}, {Hill}, \& {Kowalski}}]{dra09}
{Drake}, A.~J., {Djorgovski}, S.~G., {Mahabal}, A., {et~al.} 2009, \apj, 696,
  870

\bibitem[{{Drake} {et~al.}(2013{\natexlab{a}}){Drake}, {Catelan}, {Djorgovski},
  {Torrealba}, {Graham}, {Belokurov}, {Koposov}, {Mahabal}, {Prieto},
  {Donalek}, {Williams}, {Larson}, {Christensen}, \& {Beshore}}]{dra13}
{Drake}, A.~J., {Catelan}, M., {Djorgovski}, S.~G., {et~al.}
  2013{\natexlab{a}}, \apj, 763, 32

\bibitem[{{Drake} {et~al.}(2013{\natexlab{b}}){Drake}, {Catelan}, {Djorgovski},
  {Torrealba}, {Graham}, {Mahabal}, {Prieto}, {Donalek}, {Williams}, {Larson},
  {Christensen}, \& {Beshore}}]{dra13b}
---. 2013{\natexlab{b}}, \apj, 765, 154

\bibitem[{{Grillmair}(2006)}]{gri06}
{Grillmair}, C.~J. 2006, \apjl, 645, L37

\bibitem[{{Grillmair} \& {Dionatos}(2006)}]{gd06}
{Grillmair}, C.~J., \& {Dionatos}, O. 2006, \apjl, 643, L17

\bibitem[{{Horne}(1986)}]{hor86}
{Horne}, K. 1986, \pasp, 98, 609

\bibitem[{{Ivezi{\'c}} {et~al.}(2007){Ivezi{\'c}}, {Smith}, {Miknaitis}, {Lin},
  {Tucker}, {Lupton}, {Gunn}, {Knapp}, {Strauss}, {Sesar}, {Doi}, {Tanaka},
  {Fukugita}, {Holtzman}, {Kent}, {Yanny}, {Schlegel}, {Finkbeiner},
  {Padmanabhan}, {Rockosi}, {Juri{\'c}}, {Bond}, {Lee}, {Stoughton}, {Jester},
  {Harris}, {Harding}, {Morrison}, {Brinkmann}, {Schneider}, \& {York}}]{ive07}
{Ivezi{\'c}}, {\v Z}., {Smith}, J.~A., {Miknaitis}, G., {et~al.} 2007, \aj,
  134, 973

\bibitem[{{Koposov} {et~al.}(2010){Koposov}, {Rix}, \& {Hogg}}]{krh10}
{Koposov}, S.~E., {Rix}, H.-W., \& {Hogg}, D.~W. 2010, \apj, 712, 260

\bibitem[{{Larson}(2007)}]{lar07}
{Larson}, S. 2007, in IAU Symposium, Vol. 236, IAU Symposium, ed. G.~B.
  {Valsecchi}, D.~{Vokrouhlick{\'y}}, \& A.~{Milani}, 323--328

\bibitem[{{Law} {et~al.}(2005){Law}, {Johnston}, \& {Majewski}}]{law05}
{Law}, D.~R., {Johnston}, K.~V., \& {Majewski}, S.~R. 2005, \apj, 619, 807

\bibitem[{{Law} \& {Majewski}(2010)}]{lm10}
{Law}, D.~R., \& {Majewski}, S.~R. 2010, \apj, 714, 229

\bibitem[{{Law} {et~al.}(2009){Law}, {Kulkarni}, {Dekany}, {Ofek}, {Quimby},
  {Nugent}, {Surace}, {Grillmair}, {Bloom}, {Kasliwal}, {Bildsten}, {Brown},
  {Cenko}, {Ciardi}, {Croner}, {Djorgovski}, {van Eyken}, {Filippenko}, {Fox},
  {Gal-Yam}, {Hale}, {Hamam}, {Helou}, {Henning}, {Howell}, {Jacobsen},
  {Laher}, {Mattingly}, {McKenna}, {Pickles}, {Poznanski}, {Rahmer}, {Rau},
  {Rosing}, {Shara}, {Smith}, {Starr}, {Sullivan}, {Velur}, {Walters}, \&
  {Zolkower}}]{law09}
{Law}, N.~M., {Kulkarni}, S.~R., {Dekany}, R.~G., {et~al.} 2009, \pasp, 121,
  1395

\bibitem[{{Layden}(1994)}]{lay94}
{Layden}, A.~C. 1994, \aj, 108, 1016

\bibitem[{{Lee} {et~al.}(2008){Lee}, {Beers}, {Sivarani}, {Allende Prieto},
  {Koesterke}, {Wilhelm}, {Re Fiorentin}, {Bailer-Jones}, {Norris}, {Rockosi},
  {Yanny}, {Newberg}, {Covey}, {Zhang}, \& {Luo}}]{lee08}
{Lee}, Y.~S., {Beers}, T.~C., {Sivarani}, T., {et~al.} 2008, \aj, 136, 2022

\bibitem[{{Lee} {et~al.}(2011){Lee}, {Beers}, {Allende Prieto}, {Lai},
  {Rockosi}, {Morrison}, {Johnson}, {An}, {Sivarani}, \& {Yanny}}]{lee11}
{Lee}, Y.~S., {Beers}, T.~C., {Allende Prieto}, C., {et~al.} 2011, \aj, 141, 90

\bibitem[{{Lupton}(1993)}]{lup93}
{Lupton}, R. 1993, {Statistics in theory and practice} (Princeton, N.J.:
  Princeton University Press)

\bibitem[{{Madore} \& {Freedman}(2012)}]{mad12}
{Madore}, B.~F., \& {Freedman}, W.~L. 2012, \apj, 744, 132

\bibitem[{{Majewski} {et~al.}(2003){Majewski}, {Skrutskie}, {Weinberg}, \&
  {Ostheimer}}]{maj03}
{Majewski}, S.~R., {Skrutskie}, M.~F., {Weinberg}, M.~D., \& {Ostheimer}, J.~C.
  2003, \apj, 599, 1082

\bibitem[{{Martin} {et~al.}(2007){Martin}, {Ibata}, {Chapman}, {Irwin}, \&
  {Lewis}}]{mar08}
{Martin}, N.~F., {Ibata}, R.~A., {Chapman}, S.~C., {Irwin}, M., \& {Lewis},
  G.~F. 2007, \mnras, 380, 281

\bibitem[{{Martinez} {et~al.}(2011){Martinez}, {Minor}, {Bullock},
  {Kaplinghat}, {Simon}, \& {Geha}}]{mar11}
{Martinez}, G.~D., {Minor}, Q.~E., {Bullock}, J., {et~al.} 2011, \apj, 738, 55

\bibitem[{{Morrison} {et~al.}(2001){Morrison}, {Olszewski}, {Mateo}, {Norris},
  {Harding}, {Dohm-Palmer}, \& {Freeman}}]{mor01}
{Morrison}, H.~L., {Olszewski}, E.~W., {Mateo}, M., {et~al.} 2001, \aj, 121,
  283

\bibitem[{{Munari} {et~al.}(2005){Munari}, {Sordo}, {Castelli}, \&
  {Zwitter}}]{mun05}
{Munari}, U., {Sordo}, R., {Castelli}, F., \& {Zwitter}, T. 2005, \aap, 442,
  1127

\bibitem[{{Munn} {et~al.}(2004){Munn}, {Monet}, {Levine}, {Canzian}, {Pier},
  {Harris}, {Lupton}, {Ivezi{\'c}}, {Hindsley}, {Hennessy}, {Schneider}, \&
  {Brinkmann}}]{mun04}
{Munn}, J.~A., {Monet}, D.~G., {Levine}, S.~E., {et~al.} 2004, \aj, 127, 3034

\bibitem[{{Newberg} {et~al.}(2010){Newberg}, {Willett}, {Yanny}, \&
  {Xu}}]{new10}
{Newberg}, H.~J., {Willett}, B.~A., {Yanny}, B., \& {Xu}, Y. 2010, \apj, 711,
  32

\bibitem[{{Ofek} {et~al.}(2011){Ofek}, {Frail}, {Breslauer}, {Kulkarni},
  {Chandra}, {Gal-Yam}, {Kasliwal}, \& {Gehrels}}]{ofe11}
{Ofek}, E.~O., {Frail}, D.~A., {Breslauer}, B., {et~al.} 2011, \apj, 740, 65

\bibitem[{{Ofek} {et~al.}(2012{\natexlab{a}}){Ofek}, {Laher}, {Law}, {Surace},
  {Levitan}, {Sesar}, {Horesh}, {Poznanski}, {van Eyken}, {Kulkarni}, {Nugent},
  {Zolkower}, {Walters}, {Sullivan}, {Ag{\"u}eros}, {Bildsten}, {Bloom},
  {Cenko}, {Gal-Yam}, {Grillmair}, {Helou}, {Kasliwal}, \& {Quimby}}]{ofe12a}
{Ofek}, E.~O., {Laher}, R., {Law}, N., {et~al.} 2012{\natexlab{a}}, \pasp, 124,
  62

\bibitem[{{Ofek} {et~al.}(2012{\natexlab{b}}){Ofek}, {Laher}, {Surace},
  {Levitan}, {Sesar}, {Horesh}, {Law}, {van Eyken}, {Kulkarni}, {Prince},
  {Nugent}, {Sullivan}, {Yaron}, {Pickles}, {Ag{\"u}eros}, {Arcavi},
  {Bildsten}, {Bloom}, {Cenko}, {Gal-Yam}, {Grillmair}, {Helou}, {Kasliwal},
  {Poznanski}, \& {Quimby}}]{ofe12b}
{Ofek}, E.~O., {Laher}, R., {Surace}, J., {et~al.} 2012{\natexlab{b}}, \pasp,
  124, 854

\bibitem[{{Oke} {et~al.}(1962){Oke}, {Giver}, \& {Searle}}]{oke62}
{Oke}, J.~B., {Giver}, L.~P., \& {Searle}, L. 1962, \apj, 136, 393

\bibitem[{{Oke} \& {Gunn}(1982)}]{og82}
{Oke}, J.~B., \& {Gunn}, J.~E. 1982, \pasp, 94, 586

\bibitem[{{Rau} {et~al.}(2009){Rau}, {Kulkarni}, {Law}, {Bloom}, {Ciardi},
  {Djorgovski}, {Fox}, {Gal-Yam}, {Grillmair}, {Kasliwal}, {Nugent}, {Ofek},
  {Quimby}, {Reach}, {Shara}, {Bildsten}, {Cenko}, {Drake}, {Filippenko},
  {Helfand}, {Helou}, {Howell}, {Poznanski}, \& {Sullivan}}]{rau09}
{Rau}, A., {Kulkarni}, S.~R., {Law}, N.~M., {et~al.} 2009, \pasp, 121, 1334

\bibitem[{{Sanders} \& {Binney}(2013{\natexlab{a}})}]{sb13}
{Sanders}, J.~L., \& {Binney}, J. 2013{\natexlab{a}}, \mnras, 433, 1813

\bibitem[{{Sanders} \& {Binney}(2013{\natexlab{b}})}]{sb13b}
---. 2013{\natexlab{b}}, \mnras, 433, 1826

\bibitem[{{Schlegel} {et~al.}(1998){Schlegel}, {Finkbeiner}, \&
  {Davis}}]{SFD98}
{Schlegel}, D.~J., {Finkbeiner}, D.~P., \& {Davis}, M. 1998, \apj, 500, 525

\bibitem[{{Sesar}(2012)}]{sesar12}
{Sesar}, B. 2012, \aj, 144, 114

\bibitem[{{Sesar} {et~al.}(2011{\natexlab{a}}){Sesar}, {Juri{\'c}}, \&
  {Ivezi{\'c}}}]{sji11}
{Sesar}, B., {Juri{\'c}}, M., \& {Ivezi{\'c}}, {\v Z}. 2011{\natexlab{a}},
  \apj, 731, 4

\bibitem[{{Sesar} {et~al.}(2011{\natexlab{b}}){Sesar}, {Stuart}, {Ivezi{\'c}},
  {Morgan}, {Becker}, \& {Wo{\'z}niak}}]{ses11}
{Sesar}, B., {Stuart}, J.~S., {Ivezi{\'c}}, {\v Z}., {et~al.}
  2011{\natexlab{b}}, \aj, 142, 190

\bibitem[{{Sesar} {et~al.}(2010){Sesar}, {Ivezi{\'c}}, {Grammer}, {Morgan},
  {Becker}, {Juri{\'c}}, {De Lee}, {Annis}, {Beers}, {Fan}, {Lupton}, {Gunn},
  {Knapp}, {Jiang}, {Jester}, {Johnston}, \& {Lampeitl}}]{ses10}
{Sesar}, B., {Ivezi{\'c}}, {\v Z}., {Grammer}, S.~H., {et~al.} 2010, \apj, 708,
  717

\bibitem[{{Sesar} {et~al.}(2012){Sesar}, {Cohen}, {Levitan}, {Grillmair},
  {Juri{\'c}}, {Kirby}, {Laher}, {Ofek}, {Surace}, {Kulkarni}, \&
  {Prince}}]{ses12}
{Sesar}, B., {Cohen}, J.~G., {Levitan}, D., {et~al.} 2012, \apj, 755, 134

\bibitem[{{Sesar} {et~al.}(2013){Sesar}, {Ivezi{\'c}}, {Stuart}, {Morgan},
  {Becker}, {Sharma}, {Palaversa}, {Juri{\'c}}, {Wozniak}, \&
  {Oluseyi}}]{ses13}
{Sesar}, B., {Ivezi{\'c}}, {\v Z}., {Stuart}, J.~S., {et~al.} 2013, \aj, 146,
  21

\bibitem[{{Sirko} {et~al.}(2004){Sirko}, {Goodman}, {Knapp}, {Brinkmann},
  {Ivezi{\'c}}, {Knerr}, {Schlegel}, {Schneider}, \& {York}}]{sir04}
{Sirko}, E., {Goodman}, J., {Knapp}, G.~R., {et~al.} 2004, \aj, 127, 899

\bibitem[{{Stokes} {et~al.}(2000){Stokes}, {Evans}, {Viggh}, {Shelly}, \&
  {Pearce}}]{sto00}
{Stokes}, G.~H., {Evans}, J.~B., {Viggh}, H.~E.~M., {Shelly}, F.~C., \&
  {Pearce}, E.~C. 2000, Icarus, 148, 21

\bibitem[{{Vera-Ciro} \& {Helmi}(2013)}]{vch13}
{Vera-Ciro}, C., \& {Helmi}, A. 2013, ArXiv e-prints (arXiv:1304.4646)

\bibitem[{{Vickers} {et~al.}(2012){Vickers}, {Grebel}, \& {Huxor}}]{vic12}
{Vickers}, J.~J., {Grebel}, E.~K., \& {Huxor}, A.~P. 2012, \aj, 143, 86

\bibitem[{{Vivas} \& {Zinn}(2006)}]{vz06}
{Vivas}, A.~K., \& {Zinn}, R. 2006, \aj, 132, 714

\bibitem[{{Xue} {et~al.}(2008){Xue}, {Rix}, {Zhao}, {Re Fiorentin}, {Naab},
  {Steinmetz}, {van den Bosch}, {Beers}, {Lee}, {Bell}, {Rockosi}, {Yanny},
  {Newberg}, {Wilhelm}, {Kang}, {Smith}, \& {Schneider}}]{xue08}
{Xue}, X.~X., {Rix}, H.~W., {Zhao}, G., {et~al.} 2008, \apj, 684, 1143

\bibitem[{{Yanny} {et~al.}(2009{\natexlab{a}}){Yanny}, {Rockosi}, {Newberg},
  {Knapp}, {Adelman-McCarthy}, {Alcorn}, {Allam}, {Allende Prieto}, {An},
  {Anderson}, {Anderson}, {Bailer-Jones}, {Bastian}, {Beers}, {Bell},
  {Belokurov}, {Bizyaev}, {Blythe}, {Bochanski}, {Boroski}, {Brinchmann},
  {Brinkmann}, {Brewington}, {Carey}, {Cudworth}, {Evans}, {Evans}, {Gates},
  {G{\"a}nsicke}, {Gillespie}, {Gilmore}, {Nebot Gomez-Moran}, {Grebel},
  {Greenwell}, {Gunn}, {Jordan}, {Jordan}, {Harding}, {Harris}, {Hendry},
  {Holder}, {Ivans}, {Ivezi{\v c}}, {Jester}, {Johnson}, {Kent}, {Kleinman},
  {Kniazev}, {Krzesinski}, {Kron}, {Kuropatkin}, {Lebedeva}, {Lee}, {French
  Leger}, {L{\'e}pine}, {Levine}, {Lin}, {Long}, {Loomis}, {Lupton},
  {Malanushenko}, {Malanushenko}, {Margon}, {Martinez-Delgado}, {McGehee},
  {Monet}, {Morrison}, {Munn}, {Neilsen}, {Nitta}, {Norris}, {Oravetz}, {Owen},
  {Padmanabhan}, {Pan}, {Peterson}, {Pier}, {Platson}, {Re Fiorentin},
  {Richards}, {Rix}, {Schlegel}, {Schneider}, {Schreiber}, {Schwope}, {Sibley},
  {Simmons}, {Snedden}, {Allyn Smith}, {Stark}, {Stauffer}, {Steinmetz},
  {Stoughton}, {SubbaRao}, {Szalay}, {Szkody}, {Thakar}, {Sivarani}, {Tucker},
  {Uomoto}, {Vanden Berk}, {Vidrih}, {Wadadekar}, {Watters}, {Wilhelm}, {Wyse},
  {Yarger}, \& {Zucker}}]{yan09a}
{Yanny}, B., {Rockosi}, C., {Newberg}, H.~J., {et~al.} 2009{\natexlab{a}}, \aj,
  137, 4377

\bibitem[{{Yanny} {et~al.}(2009{\natexlab{b}}){Yanny}, {Newberg}, {Johnson},
  {Lee}, {Beers}, {Bizyaev}, {Brewington}, {Fiorentin}, {Harding},
  {Malanushenko}, {Malanushenko}, {Oravetz}, {Pan}, {Simmons}, \&
  {Snedden}}]{yan09b}
{Yanny}, B., {Newberg}, H.~J., {Johnson}, J.~A., {et~al.} 2009{\natexlab{b}},
  \apj, 700, 1282

\bibitem[{{York} {et~al.}(2000){York}, {Adelman}, {Anderson}, {Anderson},
  {Annis}, {Bahcall}, {Bakken}, {Barkhouser}, {Bastian}, {Berman}, {Boroski},
  {Bracker}, {Briegel}, {Briggs}, {Brinkmann}, {Brunner}, {Burles}, {Carey},
  {Carr}, {Castander}, {Chen}, {Colestock}, {Connolly}, {Crocker}, {Csabai},
  {Czarapata}, {Davis}, {Doi}, {Dombeck}, {Eisenstein}, {Ellman}, {Elms},
  {Evans}, {Fan}, {Federwitz}, {Fiscelli}, {Friedman}, {Frieman}, {Fukugita},
  {Gillespie}, {Gunn}, {Gurbani}, {de Haas}, {Haldeman}, {Harris}, {Hayes},
  {Heckman}, {Hennessy}, {Hindsley}, {Holm}, {Holmgren}, {Huang}, {Hull},
  {Husby}, {Ichikawa}, {Ichikawa}, {Ivezi{\'c}}, {Kent}, {Kim}, {Kinney},
  {Klaene}, {Kleinman}, {Kleinman}, {Knapp}, {Korienek}, {Kron}, {Kunszt},
  {Lamb}, {Lee}, {Leger}, {Limmongkol}, {Lindenmeyer}, {Long}, {Loomis},
  {Loveday}, {Lucinio}, {Lupton}, {MacKinnon}, {Mannery}, {Mantsch}, {Margon},
  {McGehee}, {McKay}, {Meiksin}, {Merelli}, {Monet}, {Munn}, {Narayanan},
  {Nash}, {Neilsen}, {Neswold}, {Newberg}, {Nichol}, {Nicinski}, {Nonino},
  {Okada}, {Okamura}, {Ostriker}, {Owen}, {Pauls}, {Peoples}, {Peterson},
  {Petravick}, {Pier}, {Pope}, {Pordes}, {Prosapio}, {Rechenmacher}, {Quinn},
  {Richards}, {Richmond}, {Rivetta}, {Rockosi}, {Ruthmansdorfer}, {Sandford},
  {Schlegel}, {Schneider}, {Sekiguchi}, {Sergey}, {Shimasaku}, {Siegmund},
  {Smee}, {Smith}, {Snedden}, {Stone}, {Stoughton}, {Strauss}, {Stubbs},
  {SubbaRao}, {Szalay}, {Szapudi}, {Szokoly}, {Thakar}, {Tremonti}, {Tucker},
  {Uomoto}, {Vanden Berk}, {Vogeley}, {Waddell}, {Wang}, {Watanabe},
  {Weinberg}, {Yanny}, {Yasuda}, \& {SDSS Collaboration}}]{yor00}
{York}, D.~G., {Adelman}, J., {Anderson}, Jr., J.~E., {et~al.} 2000, \aj, 120,
  1579

\bibitem[{{Zinn} \& {West}(1984)}]{zw84}
{Zinn}, R., \& {West}, M.~J. 1984, \apjs, 55, 45

\bibitem[{{Zucker} {et~al.}(2006){Zucker}, {Belokurov}, {Evans}, {Kleyna},
  {Irwin}, {Wilkinson}, {Fellhauer}, {Bramich}, {Gilmore}, {Newberg}, {Yanny},
  {Smith}, {Hewett}, {Bell}, {Rix}, {Gnedin}, {Vidrih}, {Wyse}, {Willman},
  {Grebel}, {Schneider}, {Beers}, {Kniazev}, {Barentine}, {Brewington},
  {Brinkmann}, {Harvanek}, {Kleinman}, {Krzesinski}, {Long}, {Nitta}, \&
  {Snedden}}]{zuc06}
{Zucker}, D.~B., {Belokurov}, V., {Evans}, N.~W., {et~al.} 2006, \apjl, 650,
  L41

\end{thebibliography}

\clearpage

\begin{deluxetable}{cccccccccccc}
\tabletypesize{\scriptsize}
\setlength{\tabcolsep}{0.02in}
\tablecolumns{12}
\tablewidth{0pc}
\tablecaption{Positions and Light Curve Parameters of RR Lyrae Targets\label{table-positions}}
\tablehead{
\colhead{Name} & \colhead{R.A.$^a$} & \colhead{Dec$^a$} & \colhead{Survey} &
\colhead{Helio.~distance} & \colhead{$\langle m \rangle^b$} & \colhead{rExt$^c$} &
\colhead{Amplitude} & \colhead{$m_0^d$} & \colhead{Period$^e$} & 
\colhead{${\rm HJD_0^f}$} & \colhead{Member$^g$} \\
\colhead{$ $} & \colhead{(deg)} & \colhead{(deg)} & \colhead{$ $} &
\colhead{(kpc)} & \colhead{(mag)} & \colhead{(mag)} &
\colhead{(mag)} & \colhead{(mag)} & \colhead{(d)} &
\colhead{(d)} & \colhead{$ $}
}
\startdata
RR0  & 145.622246 & 66.493010 & LINEAR & 20.5 & 17.233 & 0.366 & 0.87 & 17.11 & 0.581007 & 53428.808929 & low  \\
RR1  & 147.556760 & 60.410852 & LINEAR & 29.9 & 17.805 & 0.045 & 0.63 & 17.51 & 0.747907 & 54485.763320 & low  \\
RR2  & 142.588413 & 59.743341 & LINEAR & 23.5 & 17.362 & 0.086 & 0.81 & 16.99 & 0.535057 & 52621.904999 & low  \\
RR3  & 138.620393 & 53.065955 & LINEAR & 22.6 & 17.240 & 0.044 & 0.84 & 16.79 & 0.575657 & 54465.860645 & low  \\
RR4  & 142.596437 & 49.440867 & CSS    & 50.7 & 18.922 & 0.045 & 0.70 & 18.53 & 0.677648 & 54265.667221 & high \\
RR5  & 139.486634 & 49.043981 & CSS    & 48.6 & 18.890 & 0.042 & 0.80 & 18.39 & 0.595984 & 54508.734151 & medium \\
RR6  & 143.840446 & 47.091109 & PTF    & 54.9 & 19.081 & 0.032 & 1.02 & 18.49 & 0.530818 & 55887.972840 & high \\
RR7  & 141.771831 & 46.359489 & PTF    & 47.3 & 18.858 & 0.054 & 0.76 & 18.45 & 0.639017 & 55590.054047 & high \\
RR8  & 145.212171 & 45.450505 & LINEAR & 20.2 & 17.129 & 0.053 & 0.81 & 16.71 & 0.654049 & 52620.938984 & low  \\
RR9  & 144.271648 & 42.603354 & CSS    & 41.0 & 18.517 & 0.038 & 0.77 & 18.11 & 0.567199 & 54913.653005 & high \\
RR10 & 142.541300 & 42.570500 & CSS    & 49.7 & 18.828 & 0.042 & 0.55 & 18.48 & 0.649151 & 54157.679811 & high \\
RR11 & 144.881448 & 41.439236 & CSS    & 42.5 & 18.485 & 0.041 & 0.81 & 17.99 & 0.624166 & 56271.888900 & high \\
RR12 & 146.057798 & 40.220714 & CSS    & 39.9 & 18.395 & 0.050 & 0.62 & 18.08 & 0.711552 & 56334.821312 & high \\
RR13 & 143.482581 & 39.134007 & CSS    & 41.4 & 18.505 & 0.042 & 0.89 & 17.99 & 0.527853 & 54415.904058 & high \\
RR14 & 143.913227 & 38.853250 & CSS    & 46.1 & 18.706 & 0.043 & 0.63 & 18.36 & 0.504139 & 53789.793479 & high \\
RR15 & 146.447585 & 37.553258 & CSS    & 36.7 & 18.263 & 0.040 & 0.93 & 17.71 & 0.624026 & 54913.654037 & high \\
RR16 & 148.586324 & 37.191956 & CSS    & 41.7 & 18.529 & 0.038 & 0.64 & 18.11 & 0.573213 & 54941.722401 & high \\
RR17 & 142.909363 & 37.002696 & CSS    & 42.5 & 18.446 & 0.035 & 0.64 & 18.03 & 0.582839 & 55598.766679 & high \\
RR18 & 146.008547 & 36.265846 & CSS    & 40.9 & 18.464 & 0.033 & 0.76 & 17.99 & 0.594436 & 53789.812373 & high \\
RR19 & 146.390649 & 35.795310 & LINEAR & 30.4 & 17.943 & 0.031 & 0.44 & 17.70 & 0.755026 & 52722.727848 & high \\
RR20 & 147.457788 & 32.664206 & LINEAR & 23.2 & 17.443 & 0.043 & 0.44 & 17.26 & 0.645260 & 52651.869461 & medium \\
RR21 & 147.696468 & 31.085889 & LINEAR & 21.2 & 17.220 & 0.058 & 0.74 & 16.84 & 0.567720 & 53089.743601 & low  \\
RR22 & 151.719908 & 29.021450 & LINEAR & 23.2 & 17.475 & 0.064 & 0.50 & 17.28 & 0.610734 & 52672.850836 & medium \\
RR23 & 150.579833 & 26.598017 & LINEAR & 30.7 & 17.852 & 0.071 & 0.87 & 17.38 & 0.573755 & 53078.770191 & high \\
RR24 & 150.243511 & 25.826153 & LINEAR & 27.9 & 17.716 & 0.093 & 0.58 & 17.55 & 0.708142 & 54476.844880 & high \\
RR25 & 150.647213 & 25.247547 & LINEAR & 30.4 & 17.907 & 0.097 & 0.92 & 17.55 & 0.542891 & 54539.656204 & high \\
RR26 & 151.892507 & 24.831492 & LINEAR & 29.5 & 17.849 & 0.100 & 0.85 & 17.47 & 0.620861 & 53788.855568 & high \\
RR27 & 150.544334 & 24.257983 & LINEAR & 30.4 & 17.966 & 0.088 & 0.72 & 17.64 & 0.604737 & 54595.657970 & high \\
RR28 & 155.574387 & 20.410851 & LINEAR & 23.4 & 17.410 & 0.062 & 0.87 & 16.96 & 0.549226 & 52962.947104 & low  \\
RR29 & 153.996368 & 19.222735 & LINEAR & 26.1 & 17.599 & 0.076 & 0.54 & 17.40 & 0.645174 & 53816.785913 & high \\
RR30 & 153.698975 & 19.125864 & LINEAR & 25.7 & 17.547 & 0.096 & 0.38 & 17.42 & 0.630652 & 54149.788097 & high \\
RR31 & 154.238008 & 18.790623 & LINEAR & 23.0 & 17.332 & 0.084 & 1.02 & 16.81 & 0.508603 & 52648.880186 & high \\
RR32 & 154.824925 & 18.226018 & LINEAR & 28.7 & 17.899 & 0.088 & 0.80 & 17.55 & 0.578446 & 54084.925828 & high \\
RR33 & 154.469145 & 17.427796 & LINEAR & 27.9 & 17.802 & 0.089 & 0.73 & 17.47 & 0.575995 & 54207.717695 & high \\
RR34 & 154.295002 & 17.131504 & LINEAR & 27.5 & 17.741 & 0.082 & 0.89 & 17.27 & 0.513222 & 53706.970133 & high \\
RR35 & 156.791313 & 15.992450 & LINEAR & 26.3 & 17.543 & 0.087 & 0.79 & 17.16 & 0.592709 & 54175.771290 & high \\
RR36 & 156.409120 & 14.813483 & LINEAR & 24.0 & 17.537 & 0.112 & 0.86 & 17.18 & 0.527801 & 53844.685972 & low  \\
RR37 & 156.923144 & 13.434919 & LINEAR & 25.7 & 17.623 & 0.117 & 0.53 & 17.42 & 0.595330 & 53142.672909 & low  \\
RR38 & 156.670497 & 12.576313 & LINEAR & 25.3 & 17.583 & 0.119 & 0.57 & 17.38 & 0.637090 & 52614.946206 & low  \\
RR39 & 158.493827 &  9.235715 & LINEAR & 24.9 & 17.503 & 0.079 & 0.79 & 17.11 & 0.554073 & 53851.699888 & high \\
RR40 & 158.867253 &  8.726633 & LINEAR & 27.8 & 17.817 & 0.070 & 0.56 & 17.61 & 0.604450 & 54580.724145 & low  \\
RR41 & 159.425074 &  5.838440 & LINEAR & 25.9 & 17.552 & 0.069 & 0.70 & 17.24 & 0.670693 & 53464.791174 & low  \\
RR42 & 161.380586 &  4.863503 & LINEAR & 23.3 & 17.508 & 0.083 & 0.89 & 17.08 & 0.547546 & 53846.773788 & low  \\
RR43 & 160.996538 &  3.565153 & LINEAR & 27.0 & 17.602 & 0.114 & 0.56 & 17.43 & 0.618892 & 53710.968168 & high \\
RR44 & 160.475349 &  2.694449 & LINEAR & 19.5 & 17.111 & 0.107 & 0.55 & 16.93 & 0.580428 & 54153.853763 & low  \\
RR45 & 162.260863 &  1.083530 & LINEAR & 20.9 & 17.124 & 0.134 & 0.89 & 16.75 & 0.552394 & 53327.934367 & low  \\
RR46 & 161.045184 &  0.876656 & LINEAR & 28.9 & 17.917 & 0.125 & 0.61 & 17.70 & 0.591287 & 54535.792607 & high \\
RR47 & 161.622376 &  0.491299 & LINEAR & 20.9 & 17.235 & 0.116 & 0.89 & 16.80 & 0.463190 & 54180.766355 & high \\
RR48 & 162.518750 & -2.483103 & LINEAR & 28.7 & 17.931 & 0.114 & 0.84 & 17.56 & 0.563479 & 54552.818254 & low  \\
RR49 & 162.349340 & -2.609458 & LINEAR & 23.2 & 17.343 & 0.107 & 0.87 & 16.91 & 0.523622 & 53054.827672 & high
\enddata
\tablenotetext{a}{Equatorial J2000.0 right ascension and declination from SDSS
DR9 catalog.}
\tablenotetext{b}{Flux-averaged magnitude (corrected for interstellar medium
extinction and a magnitude offset with respect to SDSS $r$-band, as \\
$\langle m\rangle = \langle m\rangle_{not\, corrected} - {\rm rExt} - \Delta$,
see Section~\ref{targets}).}
\tablenotetext{c}{Extinction in the SDSS $r$-band calculated using the
\citet{SFD98} dust map.}
\tablenotetext{d}{Magnitude at the epoch of maximum brightness.}
\tablenotetext{e}{Period of pulsation.}
\tablenotetext{f}{Reduced Heliocentric Julian Date of maximum brightness
(HJD - 2400000).}
\tablenotetext{g}{Probability of being a member of the Orphan stream.}
\tablecomments{Tables~\ref{table-positions} and~\ref{table-results} are
available as a single table in the electronic edition of the Journal.}
\end{deluxetable}

\clearpage

\begin{deluxetable}{ccccccccc}
\tabletypesize{\scriptsize}
\setlength{\tabcolsep}{0.02in}
\tablecolumns{9}
\tablewidth{0pc}
\tablecaption{Positions and Light Curve Parameters of CSS RRab Stars without
Spectroscopic Observations\label{css_rr}}
\tablehead{
\colhead{R.A.$^a$} & \colhead{Dec$^a$} & 
\colhead{Helio.~distance$^b$} & \colhead{$\langle m \rangle^c$} &
\colhead{rExt$^d$} & \colhead{Amplitude} & \colhead{$m_0^e$} &
\colhead{Period$^f$} & \colhead{${\rm HJD_0^g}$} \\
\colhead{(deg)} & \colhead{(deg)} &
\colhead{(kpc)} & \colhead{(mag)} &
\colhead{(mag)} & \colhead{(mag)} & \colhead{(mag)} &
\colhead{(d)} & \colhead{(d)}
}
\startdata
133.171530 & 54.068280 & 38.8 & 18.412 & 0.059 & 0.67 & 18.05 & 0.596552 & 53745.681875 \\
132.773220 & 50.755860 & 35.4 & 18.217 & 0.069 & 0.95 & 17.69 & 0.652085 & 54946.735650
\enddata
\tablenotetext{a}{Equatorial J2000.0 right ascension and declination.}
\tablenotetext{b}{Calculated assuming ${\rm [Fe/H]}=-2.0$ dex and $M_{RR}=0.47$
mag.}
\tablenotetext{c}{Flux-averaged magnitude (corrected for interstellar medium
extinction and a magnitude offset with respect to SDSS $r$-band, as
$\langle m\rangle = \langle m\rangle_{not\, corrected} - {\rm rExt} - \Delta$,
see Section~\ref{targets}).}
\tablenotetext{d}{Extinction in the SDSS $r$-band calculated using the
\citet{SFD98} dust map.}
\tablenotetext{e}{Magnitude at the epoch of maximum brightness.}
\tablenotetext{f}{Period of pulsation.}
\tablenotetext{g}{Reduced Heliocentric Julian Date of maximum brightness
(HJD - 2400000).}
\tablecomments{Table~\ref{css_rr} is published in its entirety in the electronic
edition of the Journal. A portion is shown here for guidance regarding its form
and content.}
\end{deluxetable}

\clearpage

\begin{deluxetable}{cccccccccccc}
\tabletypesize{\scriptsize}
\setlength{\tabcolsep}{0.02in}
\tablecolumns{11}
\tablewidth{0pc}
\tablecaption{Line-of-sight Velocities, Proper Motions, and Metallicities\label{table-results}}
\tablehead{
\colhead{Name} & \colhead{R.A.$^a$} & \colhead{Dec$^a$} &
\colhead{Helio.~distance} & \colhead{$v_{helio}^a$} & \colhead{$v_{gsr}^b$} &
\colhead{$\mu_{l}^c$} & \colhead{$\mu_{b}^d$} &
\colhead{$\mu_{err}^e$} & \colhead{${\rm [Fe/H]}$} & \colhead{Member$^f$} \\
\colhead{$ $} & \colhead{(deg)} & \colhead{(deg)} &
\colhead{(kpc)} & \colhead{(km s$^{-1}$)} & \colhead{(km s$^{-1}$)} &
\colhead{(mas yr$^{-1}$)} & \colhead{(mas yr$^{-1}$)} &
\colhead{(mas yr$^{-1}$)} & \colhead{(dex)} & \colhead{$ $}
}
\startdata
RR0  & 145.622246 & 66.493010 & 20.5 & $-114.9 \pm 21.2$ & -22.2 & 4.82 & 1.81 & 2.68 & -1.31 & low \\
RR1  & 147.556760 & 60.410852 & 29.9 & $11.5 \pm 15.9$ & 82.9 & 9.77 & 0.78 & 3.12 & -2.39 & low \\
RR2  & 142.588413 & 59.743341 & 23.5 & $99.0 \pm 11.9$ & 166.6 & 4.22 & 0.16 & 2.84 & -2.06 & low \\
RR3  & 138.620393 & 53.065955 & 22.6 & $-46.6 \pm 13.0$ & -5.0 & 1.36 & 1.53 & 2.81 & -2.23 & low \\
RR4  & 142.596437 & 49.440867 & 50.7 & $50.0 \pm 15.3$ & 78.5 & 1.03 & 2.77 & 3.35 & -2.32 & high \\
RR5  & 139.486634 & 49.043981 & 48.6 & $-38.9 \pm 16.5$ & -12.7 & 6.64 & -1.07 & 3.57 & -2.05 & medium \\
RR6  & 143.840446 & 47.091109 & 54.9 & $94.3 \pm 14.7$ & 114.1 & -3.49 & -4.27 & 3.35 & -2.37 & high \\
RR7  & 141.771831 & 46.359489 & 47.3 & $87.7 \pm 17.4$ & 104.0 & 2.54 & -0.72 & 3.43 & -1.94 & high \\
RR8  & 145.212171 & 45.450505 & 20.2 & $-85.1 \pm 13.3$ & -71.2 & 2.07 & 0.98 & 2.68 & -1.62 & low \\
RR9  & 144.271648 & 42.603354 & 41.0 & $124.2 \pm 18.5$ & 126.7 & -3.10 & -1.34 & 3.37 & -2.08 & high \\
RR10 & 142.541300 & 42.570500 & 49.7 & $111.6 \pm 16.3$ & 113.4 & -0.06 & -5.99 & 3.50 & -2.53 & high \\
RR11 & 144.881448 & 41.439236 & 42.5 & $142.8 \pm 18.6$ & 141.0 & -0.80 & -1.83 & 3.21 & -2.56 & high \\
RR12 & 146.057798 & 40.220714 & 39.9 & $123.9 \pm 14.2$ & 117.9 & 1.49 & -2.27 & 3.19 & -2.35 & high \\
RR13 & 143.482581 & 39.134007 & 41.4 & $110.7 \pm 17.2$ & 99.4 & 1.92 & -0.04 & 3.11 & -2.22 & high \\
RR14 & 143.913227 & 38.853250 & 46.1 & $60.2 \pm 17.3$ & 48.0 & -3.02 & 2.93 & 3.15 & -2.36 & high \\
RR15 & 146.447585 & 37.553258 & 36.7 & $182.6 \pm 15.2$ & 166.4 & -4.52 & 0.01 & 3.13 & -2.14 & high \\
RR16 & 148.586324 & 37.191956 & 41.7 & $144.5 \pm 15.2$ & 128.0 & 0.23 & 3.03 & 3.22 & -2.18 & high \\
RR17 & 142.909363 & 37.002696 & 42.5 & $138.9 \pm 16.4$ & 119.2 & 1.55 & 1.25 & 2.99 & -2.73 & high \\
RR18 & 146.008547 & 36.265846 & 40.9 & $155.4 \pm 16.9$ & 134.1 & 1.79 & 4.34 & 3.23 & -2.27 & high \\
RR19 & 146.390649 & 35.795310 & 30.4 & $145.7 \pm 15.2$ & 122.7 & 7.78 & -1.57 & 3.03 & -1.96 & high \\
RR20 & 147.457788 & 32.664206 & 23.2 & $168.0 \pm 12.1$ & 133.6 & -0.51 & 1.32 & 2.72 & -1.57 & medium \\
RR21 & 147.696468 & 31.085889 & 21.2 & $-71.7 \pm 10.9$ & -112.0 & 1.04 & -3.90 & 2.63 & -1.72 & low \\
RR22 & 151.719908 & 29.021450 & 23.2 & $127.5 \pm 9.8$ & 82.0 & 2.96 & -0.87 & 2.74 & -1.46 & medium \\
RR23 & 150.579833 & 26.598017 & 30.7 & $169.3 \pm 18.0$ & 113.9 & -1.38 & -1.84 & 2.93 & -2.42 & high \\
RR24 & 150.243511 & 25.826153 & 27.9 & $207.1 \pm 13.7$ & 148.6 & 0.96 & -1.01 & 2.82 & -2.14 & high \\
RR25 & 150.647213 & 25.247547 & 30.4 & $177.7 \pm 15.0$ & 117.4 & -0.51 & -0.91 & 2.83 & -2.12 & high \\
RR26 & 151.892507 & 24.831492 & 29.5 & $145.6 \pm 15.9$ & 84.7 & 1.55 & -4.59 & 2.86 & -2.09 & high \\
RR27 & 150.544334 & 24.257983 & 30.4 & $247.3 \pm 13.9$ & 183.3 & 1.26 & 1.95 & 2.89 & -1.86 & high \\
RR28 & 155.574387 & 20.410851 & 23.4 & $-45.0 \pm 12.3$ & -118.8 & 2.64 & 1.72 & 2.85 & -1.82 & low \\
RR29 & 153.996368 & 19.222735 & 26.1 & $217.2 \pm 14.3$ & 137.7 & -3.43 & -2.89 & 2.90 & -2.00 & high \\
RR30 & 153.698975 & 19.125864 & 25.7 & $193.6 \pm 12.0$ & 113.5 & -3.91 & 2.13 & 2.83 & -2.09 & high \\
RR31 & 154.238008 & 18.790623 & 23.0 & $230.3 \pm 13.3$ & 149.5 & 1.51 & -6.12 & 2.77 & -1.97 & high \\
RR32 & 154.824925 & 18.226018 & 28.7 & $188.3 \pm 13.4$ & 106.0 & -5.80 & 1.16 & 2.94 & -1.61 & high \\
RR33 & 154.469145 & 17.427796 & 27.9 & $197.7 \pm 12.6$ & 112.3 & -0.53 & 1.41 & 2.93 & -1.75 & high \\
RR34 & 154.295002 & 17.131504 & 27.5 & $191.5 \pm 13.0$ & 104.9 & -4.38 & -6.59 & 2.77 & -1.88 & high \\
RR35 & 156.791313 & 15.992450 & 26.3 & $211.8 \pm 13.9$ & 123.6 & 0.73 & -2.51 & 2.45 & -2.32 & high \\
RR36 & 156.409120 & 14.813483 & 24.0 & $14.0 \pm 12.4$ & -78.6 & 2.07 & -3.34 & 2.43 & -1.50 & low \\
RR37 & 156.923144 & 13.434919 & 25.7 & $-23.9 \pm 13.0$ & -120.6 & 2.24 & -0.67 & 2.67 & -1.77 & low \\
RR38 & 156.670497 & 12.576313 & 25.3 & $-27.5 \pm 12.2$ & -127.4 & -2.56 & -1.54 & 2.69 & -1.80 & low \\
RR39 & 158.493827 & 9.235715 & 24.9 & $210.7 \pm 12.0$ & 101.7 & -2.66 & 2.17 & 2.51 & -2.00 & high \\
RR40 & 158.867253 & 8.726633 & 27.8 & $-64.5 \pm 14.7$ & -174.7 & 0.48 & -2.81 & 2.90 & -1.68 & low \\
RR41 & 159.425074 & 5.838440 & 25.9 & $27.7 \pm 13.2$ & -91.0 & -0.71 & -2.03 & 3.23 & -2.15 & low \\
RR42 & 161.380586 & 4.863503 & 23.3 & $-89.8 \pm 13.4$ & -209.2 & 3.92 & -1.73 & 2.80 & -1.36 & low \\
RR43 & 160.996538 & 3.565153 & 27.0 & $177.6 \pm 15.2$ & 53.8 & -2.51 & 0.56 & 3.23 & -2.31 & high \\
RR44 & 160.475349 & 2.694449 & 19.5 & $19.7 \pm 12.7$ & -107.4 & 0.17 & 0.31 & 3.00 & -1.39 & low \\
RR45 & 162.260863 & 1.083530 & 20.9 & $131.2 \pm 14.3$ & 1.5 & 1.23 & -3.83 & 3.09 & -1.97 & low \\
RR46 & 161.045184 & 0.876656 & 28.9 & $232.1 \pm 13.8$ & 100.3 & -4.70 & 0.84 & 3.27 & -1.58 & high \\
RR47 & 161.622376 & 0.491299 & 20.9 & $211.7 \pm 11.8$ & 79.5 & -1.61 & -7.65 & 2.82 & -1.50 & high \\
RR48 & 162.518750 & -2.483103 & 28.7 & $115.2 \pm 14.4$ & -24.3 & 1.54 & 1.08 & 3.31 & -1.48 & low \\
RR49 & 162.349340 & -2.609458 & 23.2 & $205.7 \pm 15.6$ & 65.6 & 2.31 & -3.29 & 2.95 & -2.02 & high
\enddata
\tablenotetext{a}{Heliocentric systemic velocity and its uncertainty.}
\tablenotetext{b}{Velocity in the Galactic standard of rest (see
Section~\ref{results}.)}
\tablenotetext{c}{Proper motion in direction of galactic longitude.}
\tablenotetext{d}{Proper motion in direction of galactic latitude.}
\tablenotetext{e}{Uncertainty in each proper motion component.}
\tablenotetext{f}{Probability of being a member of the Orphan stream.}
\tablecomments{Tables~\ref{table-positions} and~\ref{table-results} are
available as a single table in the electronic edition of the Journal.}
\end{deluxetable}

\clearpage

\begin{deluxetable}{lllllllllll}
\tabletypesize{\scriptsize} \tablecolumns{11}
\setlength{\tabcolsep}{0.02in}
\footnotesize
\tablecaption{Orphan Stream Models of \citet{new10} \label{table-models}} \tablewidth{0pt}
\tablehead{
\colhead{$N$} & \colhead{$M_{\rm Bulge}$} & \colhead{Disk} & \colhead{$M_{\rm disk}$} & \colhead{Halo} & \colhead{$M_{\rm halo}$} & \colhead{$d/r_s$} &
\colhead{$v_{\rm halo}$} & \colhead{$M^a_{60}$} & \colhead{$\chi^2\, ^b$} &
\colhead{Fits $V_c^c$} \\
\colhead{} & \colhead{$10^{10} M_\odot$} & \colhead{type} & \colhead{$10^{10} M_\odot$} & \colhead{type} & \colhead {$10^{10} M_\odot$} & \colhead{kpc} &
\colhead{$\rm km~s^{-1}$} & \colhead{$10^{10} M_\odot$} & \colhead{} & \colhead{}
}
\startdata
1 & 1.5 & Exp &  5 & NFW & 33   & 22.25 & 155         & 40   &  2.8 & no  \\
2 & 1.5 & Exp &  5 & NFW & 20   & 22.25 & $120\pm 7$  & 24   &  3.5 & no  \\
3 & 1.5 & Exp &  5 & Log & 17.6 & 12    & $81\pm 12$  & 26.5 &  3.5 & no  \\
4 & 3.4 & M-N & 10 & Log & 35   & 12    & 114         & 47   &  4.3 & yes \\
5 & 3.4 & M-N & 10 & Log & 14   & 12    & $73\pm 24$  & 26.4 &  2.3 & yes \\
6 & 3.4 & M-N & 10 & NFW & 33   & 22.25 & 155         & 43.5 & 10.0 & yes \\
7 & 3.4 & M-N & 10 & NFW & 16   & 22.25 & $109\pm 31$ & 28.4 &  2.3 & yes
\enddata
\tablenotetext{a}{Mass due to sum of bulge, disk, and halo components out to $R = 60$ kpc.}
\tablenotetext{b}{Reduced $\chi^2$ of the orbit fit to distances of likely
Orphan stream RRab stars located beyond 33 kpc.}
\tablenotetext{c}{Circular velocity of $V_c\sim220$ {\kms} \citep{krh10,bov12}
near the Sun's radius (see Figures~13 and~15 of \citealt{new10}).}
\tablecomments{A portion of this Table has been copied from Table~3 of \citet{new10}.}
\end{deluxetable}

\end{document}